\documentclass[aps,twocolumn,superscriptaddress,amsmath,amssymb,nofootinbib]{revtex4-2}

\usepackage{graphicx}
\usepackage{dcolumn}% Align table columns on decimal point
\usepackage{bm}% textbf math
\usepackage{color}
\usepackage{txfonts}
\usepackage{hyperref}
\usepackage{soul}
\usepackage{CJK}
\usepackage{comment}

\newcommand{\normw}{\tilde{w}}
\newcommand{\effk}{\tilde{k}_{\alpha}^{\rightarrow}}
\newcommand{\mdk}{\kappa_{\alpha}^{\rightarrow}}
\newcommand{\mdshnk}{\kappa^{\rightarrow}}

\begin{document}

\begin{CJK}{UTF8}{mj}

\title{Uncovering hidden dependency in weighted networks via information entropy}
\author{Mi Jin Lee (이미진)}
\affiliation{Department of Applied Physics, Hanyang University, Ansan 15588, Korea}
\author{Eun Lee (이은)}
\affiliation{BioFrontiers Institute, University of Colorado, Boulder, Colorado 80303, USA}
\author{Byunghwee Lee (이병휘)} 
\affiliation{National Science Museum, Daejeon 34141, Korea}
\affiliation{Department of Physics, Korea Advanced Institute of Science and Technology, Daejeon 34143, Korea}
\affiliation{Natural Science Research Institute, Korea Advanced Institute of Science and Technology, Daejeon 34141, Korea}
\author{Hawoong Jeong (정하웅)}
\affiliation{Department of Physics, Korea Advanced Institute of Science and Technology, Daejeon 34141, Korea}
\author{Deok-Sun Lee (이덕선)}
\affiliation{School of Computational Sciences, Korea Institute for Advanced Study, Seoul 02455, Korea}
\author{Sang Hoon Lee (이상훈)}
\email[Corresponding author: ]{lshlj82@gnu.ac.kr}
\affiliation{Department of Physics and Research Institute of Natural Science, Gyeongsang National University, Jinju 52828, Korea}
\affiliation{Future Convergence Technology Research Institute, Gyeongsang National University, Jinju 52849, Korea}

\begin{abstract}
Interactions between elements, which are usually represented by networks, have to delineate potentially unequal relationships in terms of their relative importance or direction. The intrinsic unequal relationships of such kind, however, are opaque or hidden in numerous real systems.
For instance, when a node in a network with limited interaction capacity spends its capacity to its neighboring nodes, the allocation of the total amount of interactions to them can be vastly diverse. Even if such potentially heterogeneous interactions epitomized by weighted networks are observable, as a result of the aforementioned ego-centric allocation of interactions, the relative importance or dependency between two interacting nodes can only be implicitly accessible. 
In this work, we precisely pinpoint such relative dependency by proposing the framework to discover hidden dependent relations extracted from weighted networks. 
For a given weighted network, we provide a systematic criterion to select the most essential interactions for individual nodes based on the concept of information entropy. The criterion is symbolized by assigning the effective number of neighbors or the effective out-degree to each node, and the resultant directed subnetwork decodes the hidden dependent relations by leaving only the most essential directed interactions. 
We apply our methodology to two time-stamped empirical network data, namely the international trade relations between nations in the world trade web (WTW) and the network of people in the historical record of Korea, Annals of the Joseon Dynasty (AJD). Based on the data analysis, we discover that the properties of mutual dependency encoded in the two systems are vastly different. The nations in the WTW show much more asymmetric dependent relations than its random counterpart, which implies the global economic inequality in international trades. In contrast, the relationships of people in the AJD are much more mutual than the nations in the WTW. 
The difference comes from nontrivial correlations (or lack thereof) in the networks, for which we provide the relevant network properties and representative example nations in the case of the WTW.
\end{abstract}
\date{\today}

\maketitle

\section{Introduction}
\label{sec:intro}

We observe multitudinous emergent phenomena in our surroundings beyond our expectation:
herd behaviors such as bird flocks~\cite{gill}, fish schools~\cite{fish}, and stock market bubbles~\cite{bubble}, collective intelligence~\cite{ci,wiki}, fads~\cite{fads}, and so on. 
The unexpected and intriguing phenomena stem from collective behaviors of interacting individuals in systems of interest, which is the driving motivation of statistical physics in the first place. 
In order to elucidate the origins of the phenomena, researchers naturally have paid their attention to interaction structures among the individuals. The interaction among the individuals describes their interrelationships.

One of the most popular and useful ways to understand the relationships is to employ the network representation~\cite{NetworkReview}. Each individual or constituent of a system of interest is called a node or vertex, and pairs of the nodes can be connected via so-called links or edges representing the interactions themselves. The simplest form of network is, of course, composed of binary edges, i.e., each edge exists or not. Despite its simplicity, even such a (literally) simple network representation has taught us a lot about interacting systems and their emergent phenomena, symbolized by a number of crucial concepts such as the degree (the number of neighbors of a node) distribution.
Beyond the degree from the act of simply counting the neighbors, researchers have discovered and developed more delicate metrics encoding hidden correlations inside networks for better understanding of interacting systems, such as the assortativity (basically the two-point correlation for the degree between interacting nodes)~\cite{Assort2002, Assort2003}, the clustering coefficient (the three-point correlation for the connectivity among node triplets)~\cite{clustering1, clustering2}, and even higher-order structures~\cite{Scholtes2016,Courtney2016}.
Understanding the connectivity structure is important because the structure itself can govern the resultant emergent pattern for a given dynamical rule~\cite{rmp, infectiontime}.

The aforementioned simple representation as the binary network has led us to a great deal of remarkable discoveries so far, but we have to note that simple networks utilize limited information. What we call an edge or a link in a network corresponds to a rather abstract concept of interaction, which can be vastly diverse. There are two representative ways to move on to overcome the limitation:  directed networks by taking the possible asymmetric relation ($A \to B$, but $B \not\to A$) into account and weighted networks by taking the different quantity of interactions ($A \leftrightarrow B$ versus $A \Leftrightarrow B$) into account~\cite{NetworkReview}. Imagine a mobile phone call network describing the level of directionality and intimacy between people. 
The call data contains information such as the information about identities of callers and receivers, the total number or duration of calls within a given time window, etc. 
Using the information, we can construct a directed and/or weighted network that details the social relationships much more than its binary counterpart (calling at all or not), where inevitable information loss occurs.

In particular, the directed network representation enables us to find the \emph{asymmetric} relationship between two nodes, embodied in unidirectional edges. 
In the above example, we can detect the explicit asymmetry between a node and her friend from the call log, if we obtain the log, of course. However, in the real world, there are many situations where such explicitly revealed directional relations are just out of reach for various reasons.
Then, is it possible to uncover the asymmetry or dependency between nodes hidden in networks of interest?
We can in fact generalize this process of extracting the hidden asymmetry even further, as the asymmetry is one of the plethora of intrinsic structural correlations in networks. In other words, it is deeply related to the problem of identifying the most essential interactions that govern the whole system that can happen to be asymmetric.

In spreading dynamics, for instance, it plays a crucial role as the actual substrate network. Network researchers usually assume the directed network structure to model the potential asymmetry in spreading dynamics, but the directed structure is not always transparent. For instance, in an authoritarian society, opinions of more authoritative people are highly likely to spread to less authoritative people compared to the opposite case, but the authority is usually implicitly assumed. In that case, a part of edges (directed subedges) can participate in the actual spreading dynamics of opinion as modeled in Ref.~\cite{eunlee}.
This type of hidden pathway in spreading processes on networks is extremely important in epidemic spreading, as demonstrated in the recent coronavirus disease 2019 (COVID-19) outbreak situation~\cite{Chinazzi2020,Jia2020}. In particular, the contact tracing~\cite{Kojaku2020} is reported to be one of the most effective ways to prevent the spreading, so identifying plausible directionality on top of the (undirected) contact network will add much richer information to fight this global pandemic.  

In this paper, we propose a systematic framework to extract the most meaningful relationships focused on the asymmetry between connected nodes, i.e., hidden dependency submerged in weighted networks. It consists of the process of extracting the most important neighbors for each node via the concept of the information entropy. This ego-centric viewpoint for each node naturally defines the underlying directionality. 
We take two real-world weighted networks for our analysis, one from economy and the other from history. The network of international trade between nations and the network of people in an official historical record of Korea show vastly different properties in our framework of extracting the hidden directionality. The effect of concealed asymmetry is much stronger in the former than the latter, which we detail later in regard to their other intrinsic network properties. In particular, through the hidden directionality of the international trade relations, we not only just find a hidden asymmetry, but also provide comprehensive trajectories of the changing reciprocal relation between individual nations over time. We cross-check all of the results and conclusions with the null-model networks generated from randomized weights. 

The rest of the paper is organized as follows. 
We present the procedure of extracting the asymmetric relation and a subnetwork derived from it via the information entropy in Sec.~\ref{sec:extraction}.
To evaluate the dependency of the extracted subnetwork in diverse points of view, we suggest various measures, and show the relevant results of two empirical data in Sec.~\ref{sec:results}.
We finalize the paper with further discussion is in Sec.~\ref{sec:discussion}.

\section{Extraction of directionality based on the information entropy}
\label{sec:extraction}

\begin{figure}
\includegraphics[width=0.95\columnwidth]{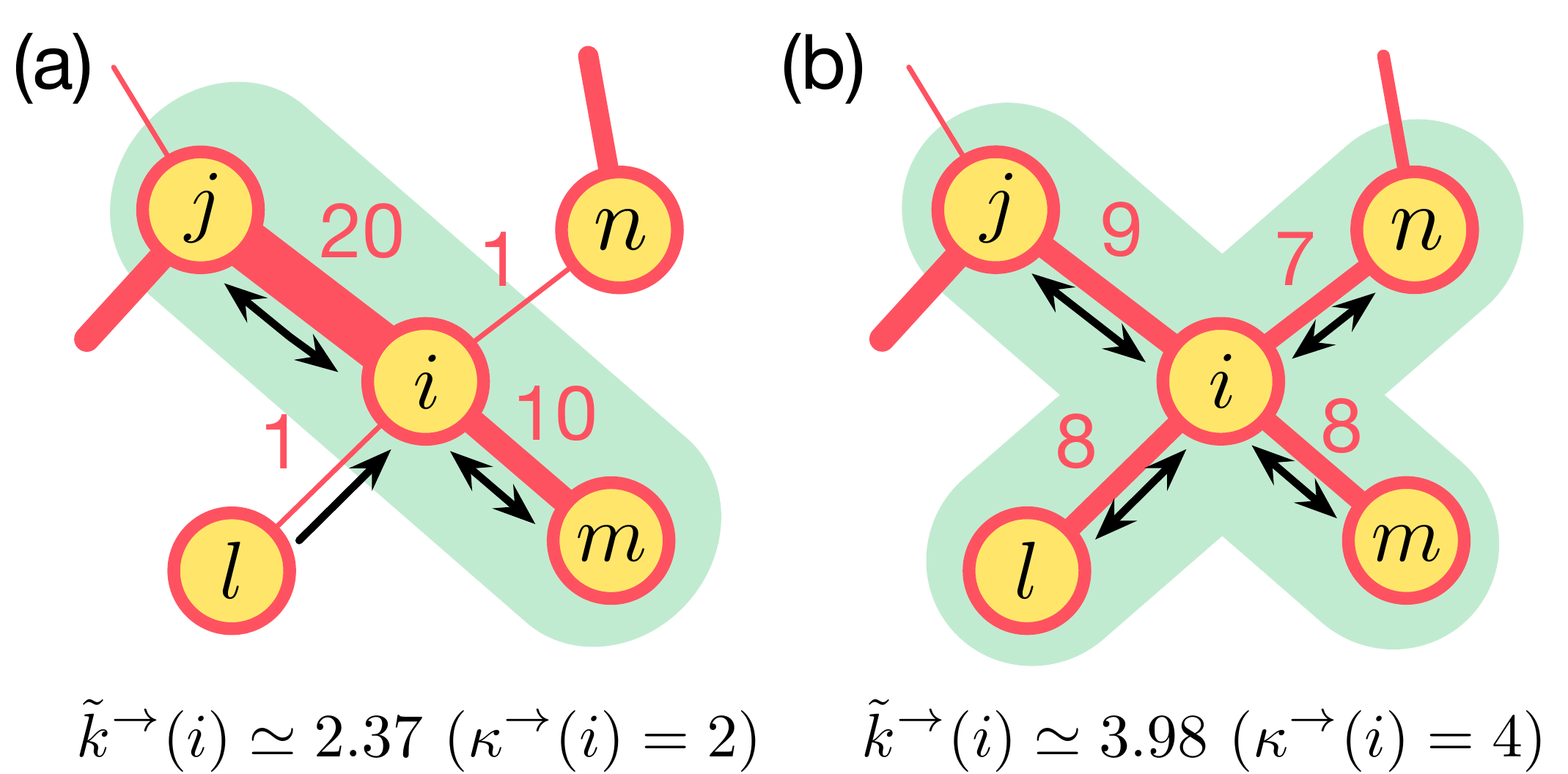}
\caption{In this example illustrating two different weight distributions, the red edges with different widths represent the original weighted networks and the black directed edges represent the resultant directed subnetwork. Each node $i$ in both (a) and (b) have the same degree $k(i)=4$ and strength $s_i = 32$, but the weights are differently distributed around each node. For node $i$ in (a), only the two neighbors with the largest and second-largest weights $\{j,m\}$ (shaded as the mint color) are chosen as the target nodes because the effective degree (with $\alpha \to 1$) $\simeq 2$ (the modified effective degree $\mdshnk=2$) while for node $i$ in (b), all of the four neighbors $\{j,l,m,n\}$ (shaded as the mint color) in the original network are chosen as the target nodes in the directed subnetwork because the effective degree (with $\alpha \to 1$) $\simeq 4$ (the modified effective degree $\mdshnk=4$).
In this hypothetical example, node $j$ keeps the edge $j \to i$, and node $n$ does not keep the edge $n \to i$ in (a), while it is the opposite case in (b).}
\label{fig:RenyiDisparityIllustration}
\end{figure}

We start to present the structural aspect of networks in which we are mainly interested.
We exemplify two representatively different cases in Fig.~\ref{fig:RenyiDisparityIllustration}. 
The node $i$ in both panels has the same degree, $4$, and the same strength (the sum of the weights on the edges connected to the node), $32$, but there is a crucial difference between node $i$ in Fig.~\ref{fig:RenyiDisparityIllustration}(a) and that in Fig.~\ref{fig:RenyiDisparityIllustration}(b), which is obviously the weight distributions around node $i$. In other words, it refers to the relative proportion of the same strength, $32$, allocated to the edges emanating from node $i$, which is essentially the cornerstone of our whole investigation.
Our main idea is that we can utilize the local or ego-centric distribution of the weights to set the quantitative criterion to pinpoint the most essential neighbors of each node, e.g., $j$ and $m$ in Fig.~\ref{fig:RenyiDisparityIllustration}(a) and all of the neighbors $j$, $l$, $m$, and $n$ in Fig.~\ref{fig:RenyiDisparityIllustration}(b), which will be shown later to be indeed the case within our framework.

To enlighten the situation a bit more deeply, take a look at the connected node pairs $(i,l)$ and $(i,m)$ in Fig.~\ref{fig:RenyiDisparityIllustration}(a).
In this example, one can easily guess that the node $i$ has two important neighbors, $j$ and $m$, and each of nodes $l$ and $m$ has only one important (in fact, the only, so indispensable) neighbor $i$. 
There is a crucial difference between the two pairs, however, because the nodes $i$ and $m$ designate each other as an important friend, while in the relation between the nodes $i$ and $l$, only node $l$ considers node $i$ an important friend and not vice versa. 
Through such asymmetry from the important friends, we can disclose the one-sided (such as $i$ and $l$) versus mutual (such as $i$ and $m$) dependency. In this section, we also present our step-by-step procedure to quantify this concept of essential neighbors and mutual importance.

\subsection{Normalized weight}
\label{subsec:normalized_weight}

Let us consider an undirected and weighted network with $N$ nodes and $L$ edges. For each node, denoted by $i$, there exists a set of weights on edges connected to its own neighboring nodes $\{ w_{ij} | j \in \nu(i) \}$ where $\nu(i)$ is the set of the neighbors of $i$, and then the cardinality of $\nu(i)$ is the number of the neighbors, or the well-known degree $k(i)$ [$k(i)=|\nu(i)|$ and $2L=\sum_{i=1}^{N}k(i)$]. The weighted adjacency matrix $\mathbf{W}$ (with its elements $w_{ij}$ for the node pair $i$ and $j$) is symmetric, i.e., $w_{ij}=w_{ji}$ where $w_{ij}>0$ if nodes $i$ and $j$ are connected, and $w_{ij}=0$ otherwise. The weight $w_{ij}$ usually represents the quantified level of interaction between $i$ and $j$, so the fraction of such interaction between $i$ and $j$ \emph{in the viewpoint of} $i$ corresponds to 
\begin{equation}
\normw_{ij} = \frac{w_{ij}}{s(i)},
\label{eq:wtilde}
\end{equation}
where $s(i) = \sum_j w_{ij}$ is called the strength of node $i$ in network terminology. 
We call the weight in Eq.~(\ref{eq:wtilde}) a \emph{normalized weight} that satisfies $\sum_j \tilde{w}_{ij} = 1$~\cite{SHLee2010}.

Let us regard the strength as the total amount of a node's resources to interact with other. 
Then the normalized weight implies how much fraction of the interaction level the node partitions to its neighbors for given limited ``resources'' of interactions.
In other words, the normalized weight $\tilde{w}_{ij}$ quantifies the importance of node $j$ from the viewpoint of node $i$.
Note $\normw_{ij} \ne \normw_{ji}$ in general even if $w_{ij} = w_{ji}$ due to the different strengths $s(i) \neq s(j) $, which is a conceptual leap presented in this work.
Accordingly, the symmetric weighted adjacency matrix $\mathbf{W}$ is cast into the asymmetric matrix $\widetilde{\mathbf{W}}$ with its element $\tilde{w}_{ij}$.
Therefore, the inequality $\tilde{w}_{ij} > \tilde{w}_{ji}$ implies that the node $j$ is more important to node $i$ than the other way around.

In the context of random walk~\cite{Noh2004} or more general types of dynamical processes~\cite{Ghavasieh2020} on networks, the normalized weight $\tilde{w}_{ij}$ represents the probability of a random walker at node $i$ to hop to an adjacent node $j$~\cite{Noh2004}. Not only the transition probability but also the actual flow of walkers may be captured by $\tilde{w}_{ij}$; In case the same number $n_\mathrm{w}$ of walkers are located at every node, the expected number of walkers $f_{ij} = n_\mathrm{w} \tilde{w}_{ij}$ hopping from node $i$ to node $j$ is proportional to $\tilde{w}_{ij}$. It will be the case for the systems and processes where every node has the same finite amount of resource for interaction, and the network with the weighted adjacency matrix $\widetilde{\mathbf{W}}$ reveals the global organization of the directed flow of walkers or information along links in them, which is not obvious but hidden in the original adjacency matrix $\mathbf{W}$. Yet we should remark that the normalized weight $\tilde{w}_{ij}$ cannot explain all types of information flow in all systems. Even in the random walk, the accumulation of such heterogeneous directed flow of walkers along edges over time eventually leads the number of walkers at a node $i$ to be proportional to its strength $s(i)$ in the stationary state such that the flow of walkers $f_{ij}$ becomes proportional to the original weight $w_{ij}$. Such a steady-state limit of random walkers also applies to the systems and processes in which every node's resource for interaction is heterogeneous or proportional to its strength, e.g., in the case of ``retweeting'' in social networking services where expansive spreading is possible. Therefore, our study based on the normalized weight in Eq.~\eqref{eq:wtilde} is limited to the systems with equal resources assigned to every node and thereby link heterogeneity emerging, such as the contact process and the transient-period random walks.

\subsection{Effective out-degree}
\label{subsec:effective_outdegree}

Based on the normalized weight defined in the previous section, we are ready to set up the scheme to extract the most essential interactions for each node.
Note that the normalized weight $\normw_{ij}$ values for node $i$ in Fig.~\ref{fig:RenyiDisparityIllustration}(a) are more heterogeneous than that in Fig.~\ref{fig:RenyiDisparityIllustration}(b).
Suppose that there are a few dominant neighbors of node $i$ whose $\normw_{ij}$ values comprise most of the interactions of node $i$ [Fig.~\ref{fig:RenyiDisparityIllustration}(a)]. 
In that case, we may suggest node $i$ to keep only those dominant neighbors and disregard the rest of less essential neighbors.
In contrast, when all of the $\normw_{ij}$ values are similar [Fig.~\ref{fig:RenyiDisparityIllustration}(b)], we can see that all of the neighbors of node $i$ are almost equally important to node $i$, so it is natural to keep all of its neighbors.
Combining the heterogeneity of $\tilde{w}_{ij}$ distribution with the fact that $\tilde{w}_{ij}$ is a probability unit, we employ the information entropy for extracting the most essential neighbors.
In Ref.~\cite{SHLee2010}, some of the authors of this paper originally introduced such a basic concept of extracting them in weighted networks, and in this paper we rigorously formulate the framework and apply it to real networks to demonstrate its utility.

The normalized weight $\normw_{ij}$ is basically a probability unit in the set $\{ \normw_{ij} | j \in \nu(i) \}$ around node $i$ (e.g., the probability of choosing $j$ out of all of the neighbors of node $i$ if $w_{ij}$ represents the unnormalized proportion of the importance of $j$ to $i$), so we employ the concept of information entropy to quantify the heterogeneity of the units allocated to each edge attached to the node. In this work, we use the R{\'e}nyi entropy~\cite{Zyczkowski2003}, which is a generalized version of information entropy with a tunable parameter to control the overall sensitivity. The R{\'e}nyi entropy~\cite{Zyczkowski2003} for node $i$ with the parameter $\alpha$ is given by
\begin{equation}
S_{\alpha} (i) = \frac{1}{1-\alpha} \ln \left( \sum_{j \in \nu(i)} \normw_{ij}^{\alpha} \right) \,.
\label{eq:renyi_disparity}
\end{equation}
The thermodynamically relevant (satisfying the additivity) Shannon entropy corresponds to the case of $\alpha \to 1$~\cite{ShannonEntropy}. 
The R{\'e}nyi entropy $S_{\alpha} (i)$ in Eq.~(\ref{eq:renyi_disparity}) approaches $\ln k(i)$ if all of the $\normw_{ij}$ values are similar, while $S_{\alpha} (i) \simeq 0$ if there exists a single neighbor $k$ dominating the interactions \emph{from} (note that we emphasize the preposition ``from'' here---we reveal its importance soon) node $i$, i.e., $\normw_{ik} \simeq 1$. Therefore, we define the \emph{effective out-degree} (again, note the prefix ``out'' and ``$\to$'' in superscript on the symbol in the following formula) of node $i$ by exponentiating $S_{\alpha} (i)$ as
\begin{equation}
\effk (i) = \exp\left[ S_{\alpha} (i) \right] = \left( \sum_{j \in \nu(i)} \normw_{ij}^{\alpha} \right)^{1/(1-\alpha)} \,,
\label{eq:effective_degree}
\end{equation}
which is also known as the Hill number~\cite{hill} to quantify a diversity of order $\alpha$ or the effective number of species in ecology~\cite{jost, Chao2016}. For the case of the Shannon entropy ($\alpha\to1$), the effective out-degree becomes
\begin{equation}
\tilde{k}^{\rightarrow}_{\alpha\to1}(i) = \exp\left[- \sum_{j \in \nu(i)} \normw_{ij} \ln \normw_{ij} \right] = \prod_{j \in \nu(i)}  \normw_{ij}^{-\normw_{ij}} \,.
\label{eq:effective_degree_shannon}
\end{equation}
In Fig.~\ref{fig:RenyiDisparityIllustration}, we provide the calculated effective out-degree values below the corresponding cases. 

As a result of exponentiating, Eq.~\eqref{eq:effective_degree} scales as $\effk (i) \simeq k(i)$ for a homogeneous weight distribution and $\effk (i) \simeq 1$ when there exists a single dominant neighbor of $i$.
Using the effective out-degree, we extract the most essential edges by taking the top $\effk (i)$ number of neighbors in the order of $\normw_{ij}$. Most importantly, those essential edges are essential in the viewpoint of $i$, so the relative importance of $\normw_{ij}$ is solely determined from $i$. This ego-centric approach naturally induces the concept of directionality, which was hidden in the original weighted network, and we detail on the core concept of this paper from it in Sec.~\ref{subsec:subnetwork}. In other words, one might take the normalized weight $\normw_{ij}$ in Eq.~\eqref{eq:wtilde} just as a local contribution of node $i$ to its neighbors, but the effective out-degree $\effk (i)$ in Eq.~(\ref{eq:effective_degree}) resulting from the nontrivially interwoven structure of the local heterogeneity in $\left\{ \normw_{ij} \right\}$ possesses the ability to extract a whole new type of information: not all of the edges, even in the case of the same weight they carry, are equally important to each of the nodes, and we can pinpoint the most significant interactions among mundane ones. %some hubs do not play a role of hubs as much as before, and small-degree nodes can act like hubs effectively around the focal nodes. 
This rather unexpected piece of information revealed by the effective out-degree is our main interest.

The effective out-degree depends not only on the heterogeneity of $\normw$ distribution but also on the parameter $\alpha$ for a given distribution of $\normw$. It is known that the R{\'e}nyi entropy is a nonincreasing function of $\alpha$ regardless of the probability distribution~\cite{Beck1993}, so as a result its exponentiated version, the effective out-degree is also non-increasing as $\alpha$ increases. In particular,
$\effk (i)=k(i)$ (it recovers the original degree regardless of the $\tilde{w}_{ij}$ distribution, except for the case $\tilde{w}_{ij}=0$ that usually corresponds to the absence of the edge between $i$ and $j$) for $\alpha=0$, whereas $\tilde{k}_{\alpha \to \infty}^{\to} (i) = 1 / \max_j \{ {\normw_{ij}} \}$, and it satisfies the inequality $1 \le \tilde{k}_{\alpha \to \infty}^{\to}(i) \le k(i)$ from $0 < \tilde{w}_{ij} \le 1$ and $|\{ w_{ij} \}| = k(i)$.
This behavior upon the parameter $\alpha$ together with the scaling behavior with respect to the heterogeneity of local weight distribution guarantees that every unisolated node has at least one essential edge in any cases. 

In particular, the case of $\alpha=2$ is widely used to quantify the heterogeneity~\cite{nowak1,nowak2,derrida,Simpson,Herfindahl,Hirschman,Serrano2009}. 
The authors of Ref.~\cite{Serrano2009} actually use $1 / \tilde{k}_{2}^{\rightarrow}$ (in our formalism) to describe the local homogeneity of weights in networks. Yet they focus on extracting the backbone structure by quantifying how peculiar the existence of each weight is compared to the null model, under the assumption of keeping the functional form of original degree distribution. As a result, their approach leads the polar opposite point to ours in the case of uniformly distributed local weights---we keep all of the neighbors because they are equally important, while Ref.~\cite{Serrano2009} does not because they are equally statistically insignificant. This is just a matter of different perspectives, and besides the fact that we use more general values of $\alpha$ in the R{\'e}nyi entropy, most importantly, we proceed one step further from here to discover the hidden directionality of weighted networks from the next section. In Appendix~\ref{sec:MBM}, we compare the results from our method to those from theirs in details for interested readers.

\subsection{Construction of a subnetwork with hidden dependency}
\label{subsec:subnetwork}

As we already introduced in Sec.~\ref{subsec:effective_outdegree}, to extract the essential neighbors from the viewpoint of each individual node, we choose only the top $\effk (i)$ neighbors, in the order of $\normw$. We illustrate the process in Fig.~\ref{fig:RenyiDisparityIllustration}.
Because the calculated effective out-degree is a real number, to practically use it (we need to ``cut'' the neighbors somewhere) we round off $\tilde{k}$ to the nearest integer $\mathcal{K}_{\alpha}^{\to} \equiv \lfloor \effk+0.5 \rfloor$.
However, a practical issue can arise if we just apply the $\mathcal{K}_{\alpha}^{\to}$ without actually looking at the $w_{ij}$ values. Suppose there exist $\zeta$ additional neighbors with the same weight as the $\mathcal{K}_{\alpha}^{\to}$-th weight in the descending order. Then, it would be unfair if we blindly take only up to the $\mathcal{K}_{\alpha}^{\to}$-th weight, because some of the neighbors with exactly the same weights are taken, and the others are not from pure luck. In that case, we decide to keep all of such neighbors. Formally, therefore,
\begin{equation}
\mdk\equiv\lfloor \effk+0.5 \rfloor + \zeta \,.
\label{eq:modified_effk}
\end{equation}
In the examples in Fig.~\ref{fig:RenyiDisparityIllustration}, $\zeta=0$ so the final integer-valued effective out-degree with $\alpha \to 1$ become $\mdk=2$ and $4$, respectively.
From now on, we refer to this particular integer version of effective out-degree $\mdk$ in Eq.~\eqref{eq:modified_effk}.

Now it is time to apply this effective out-degree from all of the nodes in a network. In other words, each individual node takes only the neighbors with the top $\mdk$ values of weight from the local weight distribution from the node. Then, we obtain the subnetwork composed of the most essential edges.
A crucial phenomenon in this procedure is that for a pair of originally connected nodes $i$ and $j$, node $j$ may belong to the top $\mdk (i)$ neighbors of node $i$, but node $i$ may not. 
In this case, the resultant subnetwork includes the \emph{unidirectional} edge $i \to j$, but not $j \to i$. This hidden directionality emerges as a result of our local threshold scheme based on information theory, which corresponds to the central theme of this paper.

Mathematically speaking, the subnetwork is represented by the asymmetric binary adjacency matrix $\tilde{\mathbf{A}}_{\alpha}$ for given $\alpha$, which gives 
\begin{equation}
\mdk (i)=\sum_{j}\tilde{A}_{\alpha,ij} \,.
\label{eq:kouteff}
\end{equation}
From the adjacency matrix, the effective in-degree coming from other nodes to node $i$ is also naturally defined as 
\begin{equation}
\kappa^{\leftarrow}_{\alpha}(i)=\sum_{j}\tilde{A}_{\alpha,ji} \,. 
\label{eq:kineff}
\end{equation}
The effective out-degree sets a local threshold assigned to every node to extract a \emph{directed} backbone structure. In contrast to the global threshold in terms of weight to obtain essential subnetworks for instance, extracting the essential edges with $\mdk$ ensures that not a single node is left out because every node has at least one effective out-degree, as discussed in Sec.~\ref{subsec:effective_outdegree}. Another popular method for backbone extraction is the maximum (or minimum, depending on the definition of the weight) spanning tree (MST)~\cite{mst} which suffers from the severe restriction of (by definition) tree structure with fixed numbers of edges (one less than the number of nodes). In addition, both the global thresholding and MST cannot extract any directional information that our method naturally yields. Compared to those conventional methods, therefore, our framework of extracting the most essential and potentially directional interactions achieves the goals of finding hidden types of information and not ignoring any nodes' local characteristics at the same time.
We use the Shannon entropy as a representative case in the remaining of this paper, so we drop the subscript $\alpha \to 1$ for all the measures from now on, e.g., $\mdshnk\equiv\kappa^{\rightarrow}_{\alpha\to 1}$.
Note that the statistical method in Ref.~\cite{Serrano2009} can also be used to yield directionality in principle, although Ref.~\cite{Serrano2009} does not actually utilize it, but as we discussed in the last paragraph of Sec.~\ref{subsec:effective_outdegree}, their point of view is different from ours.

\subsection{Mutuality from the normalized weight}
\label{sec:mutuality}

One may notice that even before extracting the directed subnetwork, the asymmetry between the normalized weights $\tilde{w}_{ij} \neq \tilde{w}_{ji}$ already insinuates the hidden directionality, which is precisely the topic of this section. The simplest measure to quantify the (a)symmetry would be to calculate the Pearson correlation between the normalized weights for opposite directions, which we call \emph{mutuality}. The mutuality $M$ is thus defined as
\begin{equation}
M \equiv \frac{\sum_{i=1}^{N}\sum_{j\in\nu(i)}\left(\normw_{ij}-\mu\right)\left(\normw_{ji}-\mu\right)}{\sum_{i=1}^{N}\sum_{j\in\nu(i)}\left(\normw_{ij}-\mu\right)^2},
\label{eq:mutuality}
\end{equation}
where $\mu = N / (2L)$ is the averaged value of $\normw_{ij}$ over all of the connected nodes pairs because each node contributes exactly unity (by the definition of normalized weights) to the total summation composed of $2L$ connected node pairs. Note that $\mu = 1 / \langle k \rangle$, where the mean degree $\langle k \rangle = 2L / N$, which we will use in the forthcoming section.
Therefore, Eq.~\eqref{eq:mutuality} can be recast as
\begin{align}
M &=\frac{\sum_{i=1}^{N}\sum_{j\in \nu(i)} \left( \normw_{ij}\normw_{ji}-\mu^2 \right)}{\sum_{i=1}^{N}\sum_{j\in \nu(i)} \left( \normw_{ij}^2-\mu^2 \right) }\nonumber \\
&= \frac{\sum_{i=1}^{N}\sum_{j\neq i}\normw_{ij}\normw_{ji}-N^2 / (2L)}{\sum_{i=1}^{N}\sum_{j\neq i}\normw_{ij}^2- N^2 / (2L) } \,,
\end{align}
which is more practical because one only needs to calculate the pairwise correlation between the normalized weights for opposite directions and the second moment of normalized weights.

The mutuality can be strongly subordinated to the underlying network structure, of course.
From the definition of normalized weights, $w_{ij}=\normw_{ij} s(i)=\normw_{ji} s(j)$ in Eq.~\eqref{eq:wtilde}, the inequality between the normalized weights $\normw_{ij}>\normw_{ji}$ is equivalent to $s(i)<s(j)$.
The strength tends to increase as the degree increases statistically \emph{if we assume the absence of intrinsic nontrivial correlations}, so $k(i)<k(j)$ under the same assumption.
Thus, one has to note that the mutuality is subject to the ``baseline'' structural network properties such as the strength-strength correlation and the degree-degree correlation called the assortativity~\cite{Assort2002, Assort2003}, so we already present the mutuality with those baseline measures.

\section{Results}
\label{sec:results}

\begin{figure}
\includegraphics[width=0.95\columnwidth]{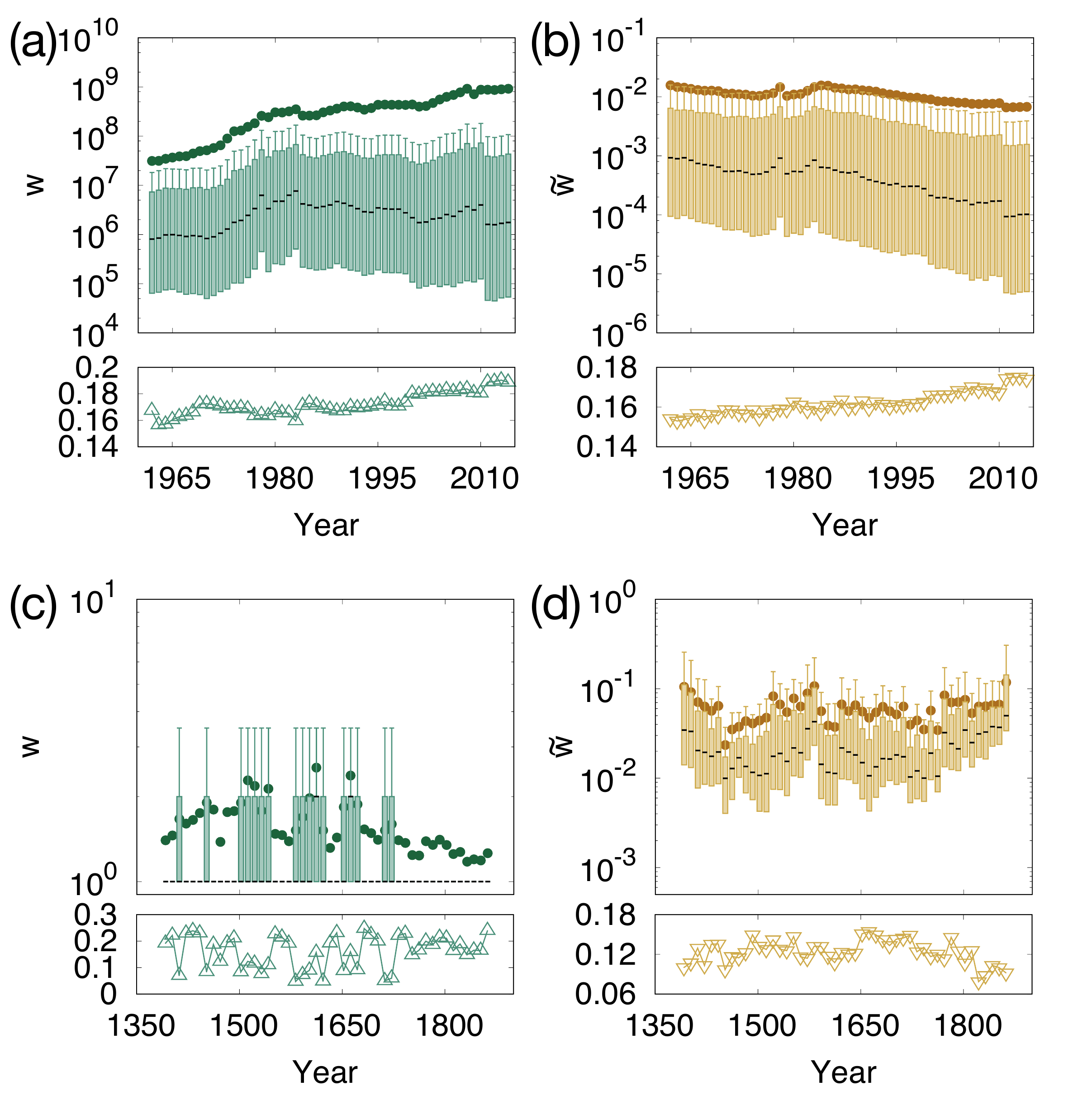}
\caption{Data description of [(a) and (b)] WTW and [(c) and (d)] AJD characterized by percentiles of the distributions of the weight $w$ and the normalized weight $\normw$. For each panel, the upper subpanel shows the boxplot of corresponding quantities along with the mean values. The vertical length of boxplot indicates the interquartile range (IQR): $Q3$--$Q1$, the black solid line represents the median ($Q2$), and the filled circles represent the mean value. The lower subpanel for each panel shows the fraction of outliers, which are defined as values $> Q3+1.5\,$IQR. As there is no value $< Q1 - 1.5\,$IQR for both data, so we only show the upper whisker indicating the outlier criterion $Q3+1.5\,$IQR. Most of $Q1, Q3$, and $Q3+1.5IQR$ in $w$ in AJD are the same as each other. Note that in the AJD, $Q1, Q3$, and $Q3+1.5IQR$ of $w$ values in (c) coincide as unity in most of the time periods.}
\label{fig:empiricaldata}
\end{figure}

\subsection{Empirical data}
\label{subsec:empricial_data}

We apply the suggested methods to two sets of empirical network data: the world trade web (WTW)~\cite{Hidalgo2007,Hidalgo2009,Hidalgo2011AAAI,WTWdata} and the Annals of the Joseon Dynasty (AJD)~\cite{AJD,AJDdata}. 
Both are time-series data between 1962 and 2014, and  1392 and 1872, respectively.
First, the WTW data is annually recorded and contains the total amount $w_{i \to j}$ of export from a nation $i$ to another nation $j$, which in turn corresponds to the total amount of import for nation $j$. We regard each nation as a node and the total amount of export as a weight on the edge from one nation to another. In other words, the WTW is orignially a directed network as $w_{i\to j}\neq w_{j\to i}$ in general. As the purpose of the current paper is to reveal the hidden directionality from originally undirected weighted networks, we intentionally construct the undirected (but weighted) version of WTW by assigning an undirected edge with the weight $w_{ij}\equiv w_{i\to j}+w_{j\to i}$ as the ``trade volume'' between two nations. 
The AJD network data is composed of the relationships between people appearing in a collection of records for historical events in Joseon Dynasty, which is a Korean dynastic kingdom that lasted for approximately five centuries (1392--1897). The network is basically a cooccurrence network, where two people are connected with the weight corresponding to the number of sentences mentioning them together within a ten-year time window. We describe more details in Appendix~\ref{sec:app_description}.

We select the WTW and AJD network data as representative examples that enable us to investigate the temporal evolution of congeneric data. First, let us brief on the most basic constituents of these weighted networks: the distribution of weights themselves and their normalized version. The time-stamped distributions of weight and the normalized weight defined in Sec.~\ref{subsec:normalized_weight} are shown in Fig.~\ref{fig:empiricaldata} (for the readers interested in more basic network measures, we show the degree and strength distributions in Appendix~\ref{sec:app_description}). Due to the heavy-tailed nature, we show the distributions by means of percentiles as the lower quartile $Q1$, the median $Q2$, and the upper quartile $Q3$, as well as the mean value. 
Both data show right-skewed distributions, reflected by the large fraction of outliers (the criterion of outliers is defined in the caption of Fig.~\ref{fig:empiricaldata}) and the fact that the mean values are always larger than the upper quartiles except for $w$ distributions in the AJD, let alone the median. 
The distributions of $w$ and $\normw$ of WTW are broader and more skewed [Figs.~\ref{fig:empiricaldata}(a) and~\ref{fig:empiricaldata}(b)] than those of AJD [Figs.~\ref{fig:empiricaldata}(c) and~\ref{fig:empiricaldata}(d)], supported by the large deviation of means from medians.

As one can clearly see from Fig.~\ref{fig:empiricaldata}, the temporal change of the normalized weight $\normw(t)$ looks almost independent of that of the original weight $w(t)$. The temporally decreasing tendency of $\normw(t)$ in WTW and fluctuating behavior of that in AJD are determined by the mean degree $\langle k\rangle = 2L / N$ because the mean value of normalized weight $\mu = N / (2L) = 1/\langle k \rangle$ (as presented in Sec.~\ref{sec:mutuality}), and the reciprocal relation is visible if one compares Figs.~\ref{fig:empiricaldata}(b) and \ref{fig:empiricaldata}(d) with Figs.~\ref{fig:meaneffek}(b) and \ref{fig:meaneffek}(d).
In addition, in the AJD, the distribution of $\normw$ looks more heterogeneous than that of $w$. We believe that a particular characteristic of these data is responsible for it; most $w$ values are concentrated on $1$ (i.e., most pairs of people appear only once in the 10-year time windows of AJD: around $80\%$ throughout the entire period) [Fig.~\ref{fig:empiricaldata}(c)], but its normalized version $\normw$ is split into different values $\normw_{ij}=1/s(i)$ and $\normw_{ji} = 1/s(j)$ from various values in $\{ s(i) \}$. 
Most of all, the overall or averaged-out distributions of $w$ and $\normw$ investigated at a global (network) level do not offer the hidden directional information we would like to discover, so let us move on to the local distribution in the next section, from which we present our core results. 

\subsection{Local distribution of the normalized weight and effective out-degree}
\label{subsec:result_effk}

\begin{figure}
\includegraphics[width=\columnwidth]{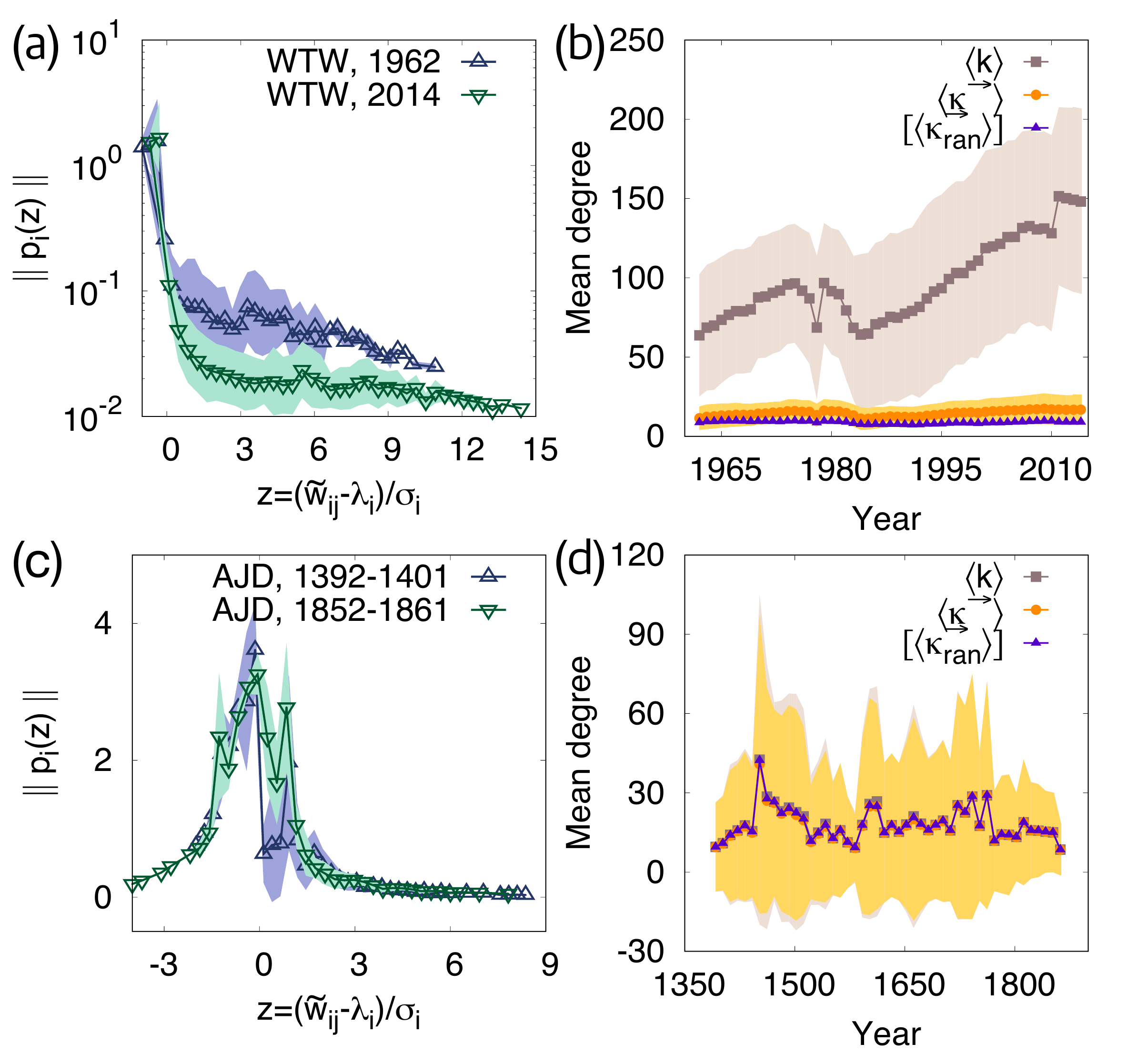}
\caption{The average distributions of normalized weights and the mean effective out-degrees based on the R{\'e}nyi entropy (the Shannon entropy in this case, as we take the $\alpha \to 1$ case). 
In the left panels, we show the average curves of the probability density function $\left \| p_i(z) \right \|$ of the rescaled normalized weight $z = (\normw_{ij} - \lambda_i)/\sigma_{i}$ defined in the main text with the standard error of $\left \| p_i(z) \right \|$ represented by the shaded area, for (a) WTW and (c) AJD, snapshots of which (one from an early period and the other from a late period) are shown. In the right panels, we show the mean values of the original degree $\langle k\rangle$, the effective out-degree $\langle \mdshnk \rangle$, and their standard deviations marked with the shaded area for (b) WTW and (d) AJD.
For comparison, we plot the mean values of effective out-degrees from 100 null-model networks with shuffled weights, denoted by $\left[\langle\mdshnk_{\rm ran}\rangle\right]$ with the standard error marked with the shaded area.}
\label{fig:meaneffek}
\end{figure}

In Sec.~\ref{subsec:effective_outdegree}, we have introduced the concept that the local distribution of the normalized weight around node $i$ denoted by $p_i(\normw_{ij})$ yields how many neighbors, i.e., $\mdk(i)$ neighbors defined as Eq.~\eqref{eq:modified_effk}, of $i$ are essential among the total $k(i)$ number of neighbors. In other words, the distribution $p_i(\normw_{ij})$ determines the effective out-degree $\mdk(i)$, so the overall shape of $p_i(\normw_{ij})$ in a network provides an informative clue to predict the $\mdk$ distribution. First, we observe that the two networks show remarkably different distributions of normalized weights. As the degree, which determines the overall scale of $\normw_{ij}$ for each $i$, is inherently heterogeneous~\cite{NetworkReview}, we have to rescale $\normw_{ij}$ first for the overview in an entire network.
The left panels of Fig.~\ref{fig:meaneffek} illustrate the representative distribution $\left \| p_i(z) \right \|$ for each data, where $z = (\normw_{ij} - \lambda_i)/\sigma_{i}$ is the rescaled variable with respect to the mean $\lambda_i = \sum_{j} \normw_{ij} / k(i) = 1 / k(i)$ and the standard deviation $\sigma_{i}$, by averaging the nonzero values of $p_i(z)$ over all of the nodes. 
In other words, for each $z$ value, the normalized weight distributions in Figs.~\ref{fig:meaneffek}(a) and \ref{fig:meaneffek}(c) are given by $\left \| p_i(z) \right \| =  \left[ {\sum_i}^{\prime} p_i(z) \right] / N(z)$ where $\sum ^{\prime}$ is the restricted sum for nonzero $p_i(z)$ values, the total number $N(z)$ of which is the normalization factor.
The normalized weight distribution of WTW is a typical heavy-tailed distribution observed in many complex interacting systems, while the distribution for AJD is unimodal and well-characterized by its mean $\lambda_i$ and standard deviation $\sigma_i$. This contrast indicates that the local distribution of weights around individual nodes in WTW is usually much more heterogeneous than that in AJD, as in the situations described in Fig.~\ref{fig:RenyiDisparityIllustration}(a) versus \ref{fig:RenyiDisparityIllustration}(b), respectively. 

Therefore, one can expect that $\mdk(i)$ of most nodes in the WTW will be smaller than their original degree $k(i)$, while most nodes in the AJD will recover their original degrees as the effective out-degrees.
The right panels of Fig.~\ref{fig:meaneffek} confirm such distinct scales of effective out-degrees with respect to the original degrees. The effective out-degrees in the WTW is much more smaller than the original degrees on average [Fig.~\ref{fig:meaneffek}(b)], while they are almost indistinguishable for the AJD [Fig.~\ref{fig:meaneffek}(d)]. More specifically, in the WTW even though the number of trading partners of nations usually increases and sometimes fluctuates in time, most nations keep a few important trading partners throughout the period. On the other hand, in the AJD, the average effective out-degree and the average original degree are almost indiscernible throughout the five centuries of Joseon Dynasty. This result verifies the expectation drawn from the local distribution of $\normw$ in the left panels of Fig.~\ref{fig:meaneffek} that there are disproportionately small numbers of essential neighbors compared to the original degree in the WTW and most neighbors are similarly important (so they are all essential according to our framework) in the AJD.

To investigate the implication of the normalized-weight distribution and the effective out-degrees in the two data further, we generate 100 null-model networks by shuffling the weight $w_{ij}$ in original networks (redistributing the weights uniformly at random to all of the existing edges) and then extract the essential edges according to the procedure described in Sec.~\ref{sec:extraction}. This shuffling process preserves the degree $k(i)$ for every node but randomizes everything related to the weight information including the original weight $w_{ij}$, the strength $s(i)$, and the normalized weight $\normw_{ij}$ for all of the nodes.
We measure the mean effective out-degree of the null-model networks, computed as $\left[\langle\mdshnk_{\rm ran}\rangle\right]$ that denotes the mean effective out-degrees for each null-model network, which are in turn averaged over the 100 null-model networks. 
As one can clearly see from Fig.~\ref{fig:meaneffek}, the AJD shows no noticeable difference between $\langle\mdshnk\rangle$ and $\left[\langle\mdshnk_{\rm ran}\rangle\right]$, while they are systematically different ($\left[\langle\mdshnk_{\rm ran}\rangle\right]$ is always smaller than $\langle\mdshnk\rangle$) in the WTW. Again, shuffling the relatively homogeneous normalized-weight distribution of the AJD does not affect the effective out-degrees of the nodes in the AJD much, because the nodes will retrieve most of their original neighbors anyway. In contrast, the fact that the effective out-degrees of randomized version of the WTW are systematically smaller than the real effective out-degrees indicates, as discussed in Fig.~\ref{fig:RenyiDisparityIllustration}, that the heterogeneity of link weights around a node is weaker in the real WTW than in the randomized WTW. The shuffling process wipes out any correlation of the link weights around a node and equate the local heterogeneity of link weights with the global-level heterogeneity delineated in Fig.~\ref{fig:empiricaldata}(a).

\begin{figure*}
\includegraphics[width=0.95\textwidth]{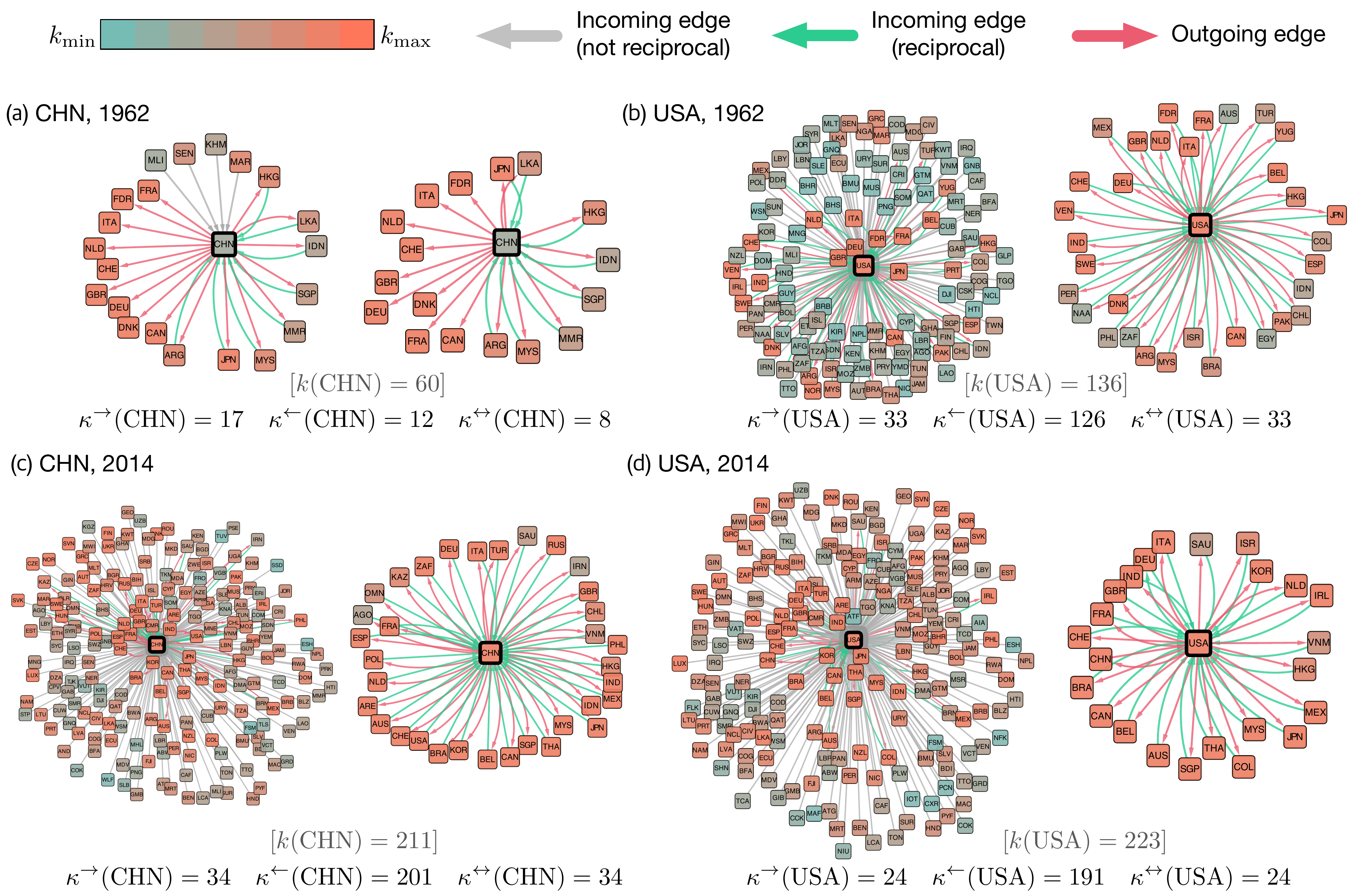}
\caption{The parts of subnetworks of WTW in [(a) and (b)] 1962 and [(c) and (d)] 2014, in the cases of [(a) and (c)] CHN- and [(b) and (d)] USA-centric viewpoint (only their adjacent neighbors with at least a type of edge in the subnetwork are shown), so CHN and USA are at the center of each panel. Nodes are colored by their original degree values $k(i)$. In each panel, a subnetwork in the left is the entire corresponding nation-centric network, where the directed edges are classified as outgoing (orange), reciprocal incoming (mint), and non-reciprocal incoming (gray) ones. On the right in each panel, only the outgoing part (ignoring the non-reciprocal incoming edges) is shown.}
\label{fig:network}
\end{figure*}

\subsection{Evaluation of dependency}
\label{subsec:dependency}

So far, we have investigated the hidden directionality by observing the averaged quantities of the most elementary measures. In this section, we take a step further into the systems of interest and suggest a few derivative measures in both global and local levels, to demonstrate the utility of our framework. 
As illustrative examples, we show parts of the subnetworks constructed by the procedure in Sec.~\ref{subsec:subnetwork}, from the oldest [Figs.~\ref{fig:network}(a) and \ref{fig:network}(b)] and latest [Figs.~\ref{fig:network}(c) and \ref{fig:network}(d)] WTW data (with nontrivial hidden dependency as revealed in previous sections); in particular, we take ego-centric view of the subnetwork from two characteristic nations, which are China (CHN) [Figs.~\ref{fig:network}(a) and~\ref{fig:network}(c)] and the United States of America (USA) [Figs.~\ref{fig:network}(b) and~\ref{fig:network}(d)].
One can see the $\mdshnk$ outgoing edges (pink) and the $\kappa^{\leftarrow}$ incoming edges (gray or green, depending on the reciprocity detailed soon), as defined in Eqs.~\eqref{eq:kouteff} and \eqref{eq:kineff}, respectively.
The outgoing and incoming edges here refer to the \emph{interaction to} trading partner nations that a nation considers essentially important and the \emph{interaction from} trading partner nations that considers the nation as such, respectively. The intersection of outgoing and incoming edges corresponds to the reciprocal edges (green) that represent the mutually important relations. The effective reciprocal degree $\kappa^{\leftrightarrow}(i) = \sum_{j} \tilde{A}_{ij}\tilde{A}_{ji}$ denotes the number of the reciprocal edges attached to node $i$, where $\tilde{A}_{ij}$ is an element of the asymmetric binary adjacency matrix in Sec.~\ref{subsec:subnetwork}. 

Not surprisingly, the enormous growth of the Chinese economy is reflected in the growth in the number of trading partner nations of China ($60 \to 211)$ over the decades between 1962 and 2014. In particular, compared to its doubled effective out-degree growth ($17 \to 34$), its effective in-degree has been increased by more than $15$ times ($12 \to 201$). As the latter indicates other nations' \emph{dependency on} China, the dramatic change in $\kappa^{\leftarrow}$ captures each nation's genuine influence to the global economy more accurately than the change in the number of trading partner nations (the original degree). In the case of USA, as expected, it was already one of the most influential nations in 1962 already and is still the case, and the numbers of its trading partner nations and its effective in-degree are increased supposedly due to the overall economic growth globally. However, at the same time, the effective out-degree of USA has been decreased ($33 \to 24$) during the period. In other words, despite the global economic growth, the international trade of USA has become more heterogeneous among its trading partner nations, which may suggest the global economic inequality. Again, we would like to emphasize that this type of distinct analyses is not possible if we only look at the conventional network measures such as degree, strength, and weight distribution without taking the hidden dependency into account.

To characterize the properties of directed subnetworks from effective out-degrees in more details,
we calculate the measures called the relative edge density $e$ and the reciprocity $r$, defined as
\begin{align}
e &\equiv \frac{\sum_{i=1}^{N}\mdshnk(i)}{\sum_{i=1}^{N}k(i)}, \label{eq:edgedensity}\\
r &\equiv \frac{\sum_{i=1}^N \kappa^{\leftrightarrow}(i)}{\sum_{i=1}^{N}\mdshnk(i)},  \label{eq:reciprocity}
\end{align}
respectively.
The relative edge density $e$ indicates the fraction of essential neighbors for the nodes in a network on average, or the homogeneity of the local weight distribution. Dividing both the numerator and the denominator in Eq.~\eqref{eq:edgedensity} by $N$, the effective edge density can be rewritten as $e = \langle\mdshnk\rangle / \langle k\rangle$, or the ratio of the mean effective out-degree to the mean degree in the right panels of Fig.~\ref{fig:meaneffek}.
The reciprocity $r$ is the fraction of the bidirectional edges among the essential edges.
It quantifies the fraction of edges in a weighted network representing the mutually (essentially) dependent relation. 

\begin{figure}
\includegraphics[width=\columnwidth]{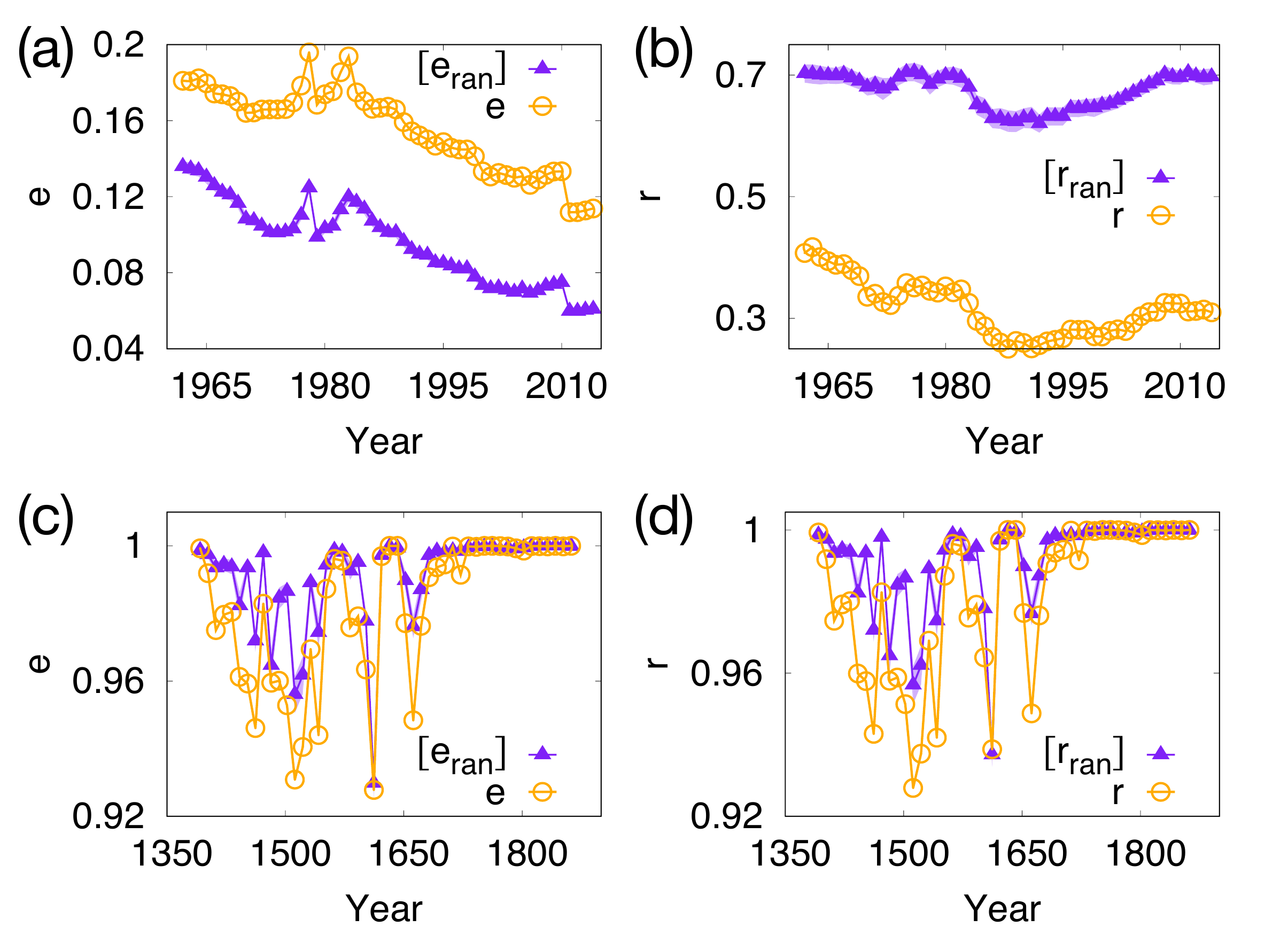}
\caption{The temporal changes of relative edge density $e$ and reciprocity $r$ are shown for [(a) and (b)] the WTW and [(c) and (d)] the AJD.
On the left panels, we plot the relative edge density $e$ for the real data and $[e_{\rm ran}]$ for the weight-shuffled null models.
On the right panels, we plot the reciprocity $r$ for the real data and $[r_{\rm ran}]$ for the weight-shuffled null models.}
\label{fig:reciprocity}
\end{figure}

We show the temporal changes of $e$ and $r$ for the WTW in the upper panels of Fig.~\ref{fig:reciprocity} and for the AJD in the lower panels of Fig.~\ref{fig:reciprocity}.
For comparison, we also plot the corresponding measures obtained from the weight-shuffled null-model networks introduced in Sec.~\ref{subsec:result_effk}.
The relative edge density $e$ in WTW stays at quite a low level roughly between $0.1$ and $0.2$ [Fig.~\ref{fig:reciprocity}(a)] with a decreasing trend over time, and the reciprocity $r$ stays around $0.3$ [Fig.~\ref{fig:reciprocity}(b)]. As already mentioned in the previous paragraph, the decreasing values of $e$ is equivalent to the overall increasing trend of $\langle k \rangle$ and the relatively flat $\langle\mdshnk\rangle$ shown in Fig.~\ref{fig:meaneffek}(b). We can interpret this in such a way that nations take part in the international trade more and more as the world trade expands as time goes by (increasing $\langle k \rangle$ over time), but their lion's share of trade is usually dominated by a few number of trading partner nations (the relatively flat $\langle \mdshnk \rangle$ over time), yielding the decreasing trend of $e$.

We clarify the implication of $e$ and $r$ by comparing them with those from the null-model networks. The relative edge density $e$ is larger than that from the null-model networks denoted by $[e_{\rm ran}]$, but the reciprocity $r$ is smaller than that from the null-model networks denoted by $[r_{\rm ran}]$, as shown in Fig.~\ref{fig:reciprocity} (again, $[ \cdots ]$ indicates the ensemble-averaged quantity). The former is expected because $\left[\langle\mdshnk_{\rm ran}\rangle\right] < \langle\mdshnk\rangle$ (Fig.~\ref{fig:meaneffek}) and $e = \langle\mdshnk\rangle / \langle k\rangle$, so $[e_{\rm ran}] < e$.
The latter ($r<[r_{\rm ran}]$) indicates that the mutually essential reactions happen less likely than a chance. The rank of the weight of a link---trade volume---may be high enough to be counted as effective for one end node but may not for the other, probably caused by the severe disparity in their overall link weights related to the national economic scales. In the randomized version, on the contrary, the links of every node are assigned weights randomly on equal foots, except for statistical fluctuation, and therefore a link assigned a high weight is likely to be counted as effective for both end nodes.

In contrast, as we have already repeatedly checked, the AJD recovers most of its original interactions as essential ones, i.e., $e \simeq 1$, as shown in Fig.~\ref{fig:reciprocity}(c), which is consistent with the result $\langle\mdshnk\rangle \simeq \langle k \rangle$ in Fig.~\ref{fig:meaneffek}(d). Moreover, the property that most original interactions are recovered in the subnetwork also means that interactions are retrieved in both directions, so the reciprocity $r \simeq 1$ as well, as shown in Fig.~\ref{fig:reciprocity}(d). Simply put, the weights in AJD do not play any significant role due to their near uniformity, which is also confirmed by the observation that the average relative edge density $[e_{\rm ran}]$ and the average reciprocity $[r_{\rm ran}]$ of their null-model networks are quite similar to $e$ and $r$ from the real AJD network, as shown in Figs.~\ref{fig:reciprocity}(c) and \ref{fig:reciprocity}(d). In other words, the weight-shuffling process does not affect the properties of AJD notably, as long as the substrate (binary) network is preserved.

\begin{figure*}
\includegraphics[width=0.95\textwidth]{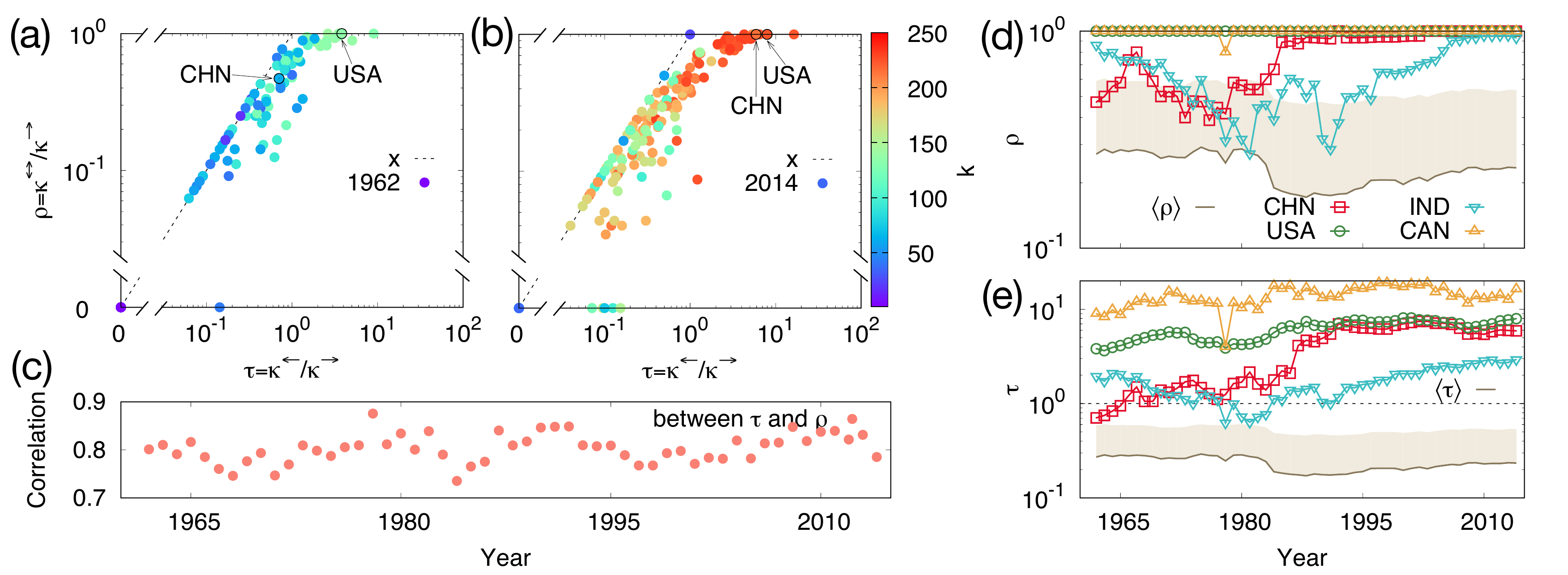}
\caption{Individual-node level LR and AR measures in WTW. [(a) and (b)] Scatter plots in the AR--LR space, where individual nations are the points, in 1962 and 2014. The nodes are colored with their original degree values. The black dashed lines indicate $\rho=\tau$, which is the upper bound of $\rho$ for a given $\tau$ as in Eq.~\eqref{eq:tau_inequality}.
(c) The temporal change of Pearson correlation coefficient between $\rho$ and $\tau$, with the $p$-values $<10^{-27}$ throughout the period.  
[(d) and (e)] The temporal changes of $\rho$ and $\tau$ for four representative nations mentioned in the main text. The solid curves represent the average values of $\langle\rho\rangle$ and $\langle\tau\rangle$ (note that $\langle \rho \rangle \neq r$ because $\langle x/y \rangle \neq \langle x \rangle / \langle y \rangle$), where the shaded upper areas indicate the standard deviations $\sigma_{\rho}$ and $\sigma_{\tau}$ (we only show the upper areas because $\langle\rho\rangle-\sigma_{\rho}<0$ and $\langle\tau\rangle-\sigma_{\tau}<0$). The horizontal dotted line $\tau = 1$ in (e) indicates the baseline AR with $\kappa^{\leftarrow}(i) = \mdshnk(i)$.
}
\label{fig:local_dependency}
\end{figure*}

From now on, we apply the concept of the reciprocity learned from the global-level analysis back to the individual-node level, where all of our framework begins in fact. As the ``global'' version of the reciprocity in Eq.~\eqref{eq:reciprocity} is from the averaged measures, we can define its ``local'' version as  
\begin{equation}
\rho(i) \equiv\frac{\kappa^{\leftrightarrow}(i)}{\mdshnk(i)} \,,
\label{eq:rho}
\end{equation}
which we call the local reciprocity (LR), and it represents how many of essential neighbors of node $i$ also consider node $i$ as their essential neighbor. There is one more thing we introduce as another meaningful measure in the local level, corresponding to the ratio of the effective in-degree to the effective out-degree as
\begin{equation}
\tau(i) \equiv \frac{\kappa^{\leftarrow}(i)}{\mdshnk(i)} \,,
\label{eq:tau}
\end{equation}
which we call the attraction ratio (AR) and describes how attractive node $i$ is to its neighbors, relative to the number of attractive neighbors to node $i$. Note that there is no global measure corresponding to AR as $\sum_i \kappa^{\leftarrow}(i) = \sum_i \mdshnk(i)$ trivially, and they always satisfy the inequalities
\begin{align}
0 \le \rho(i) \le \min[1,\tau(i)] \,, \label{eq:rho_inequality} \\
\rho(i) \le \tau(i) \le k(i)/\mdshnk(i) \label{eq:tau_inequality} \,.
\end{align}

We solely focus on the WTW data here, as not surprisingly for most nodes in the AJD network  $\rho(i)\simeq 1$ and $\tau(i)\simeq 1$.
We show the scatter plot of the local measures defined above from the WTW network in Figs.~\ref{fig:local_dependency}(a) (1962) and \ref{fig:local_dependency}(b) (2014), where each point represents each nation, and one can easily check the inequality in Eq.~\eqref{eq:rho_inequality}.
The $\rho$ and $\tau$ for CHN and USA depicted in Fig.~\ref{fig:network} are highlighted by the black empty circles and the arrows.
From the scatter plot where the nodes are color-coded with their original degree, one can recognize that nations with many trading partners tend to have large values of $\rho$ and $\tau$, and $\rho$ and $\tau$ are positively correlated [Fig.~\ref{fig:local_dependency}(c)] partly because of the upper bound of $\rho$ for given $\tau$ values in Eqs.~\eqref{eq:rho_inequality} and \eqref{eq:tau_inequality}, we suppose. Naturally, larger values of $\tau$ increase the chance for the corresponding trading partner nations that consider the nation as an essential partner to be reciprocal. The correlation is significant throughout the entire period of the data we have examined, as shown in Fig.~\ref{fig:local_dependency}(c).

The locations of nations in this $\rho$--$\tau$ space throughout the time provide an overview of the nations' status in the international trade in terms of their mutual importance to other nations. We take four nations in particular to demonstrate it: CHN, USA, India (IND), and Canada (CAN) and show their temporal changes of LR and AR in Figs.~\ref{fig:local_dependency}(d) and \ref{fig:local_dependency}(e), respectively.
As we have checked in Fig.~\ref{fig:network}, USA has maintained its theoretically maximum level of LR ($\rho=1$: all of USA's essential nations take USA as an essential trading partner all the time) throughout the entire period of the data and its status of the ``attractive'' ($\tau > 1$) trading partner to other nations with an increasing trend from $\tau \simeq 3.8$ to $\tau \simeq 8.0$.
In the case of CHN, as we have observed in Fig.~\ref{fig:network}, both AR and LR have been significantly increased for the past few decades, signifying its dramatic economic growth during the period, and one can check that the effect is more substantial for AR (proportional to the number of nations that take China as an important partner).

In particular, the AR seems to augment the distinction between the trading relations in the case of similar values of the LR; For instance, both USA and CAN maintain $\rho=1$ (except for the small dip in 1979 for CAN), but the AR for CAN is significantly larger than that for USA throughout the period, i.e., CAN is a much more ``attractive'' trading partner than USA, relative to the number of nations they respectively take seriously. Taking the different baseline values into account, the temporal trends of AR for the two nations are similar, as they belong to the same geopolitical economic block such as the North American Free Trade Agreement (NAFTA) with the large bilateral trade volume~\cite{worldbank}. 

Another characteristic nation is IND, which shows a decreasing (up to 1980s) and then increasing trend for both LR and AR measures consistent with its recent history of industrial growth~\cite{india1}. The large rearrangement in the overall international trade in the early 1980s is in fact also observed in the structural change itself, e.g, the connectivity significantly shrank, as shown in Figs.~\ref{fig:meaneffek}(b) and \ref{fig:reciprocity}(a), which may explain the small dip in CAN as well. The second oil shock~\cite{oilshock} occurred during this time may be responsible for this overall reorganization of WTW. In other words, the overall trading capacity was temporarily lowered. Particularly, IND suffered from the reduction of the overall trade volume by the international debit crisis~\cite{india2}

Albeit anecdotally, these examples demonstrate that our method of extracting the hidden dependency provides a unique viewpoint on intricate networked systems. We expect that the effect of the current COVID-19 outbreak on the international trade and global economy can also be analyzed with this type of dependency analysis in the future. Finally, we would like to remark that the weight-shuffled version of the null-model networks is used in this section, but one can try different levels of null models, e.g., synthetic model networks for baseline properties of the various measures, as we demonstrate in Appendix~\ref{sec:app_synthetic}.

\subsection{Inference to originally directed networks}
\label{subsec:inference_to_original_network}

\begin{figure}
\includegraphics[width=0.8\columnwidth]{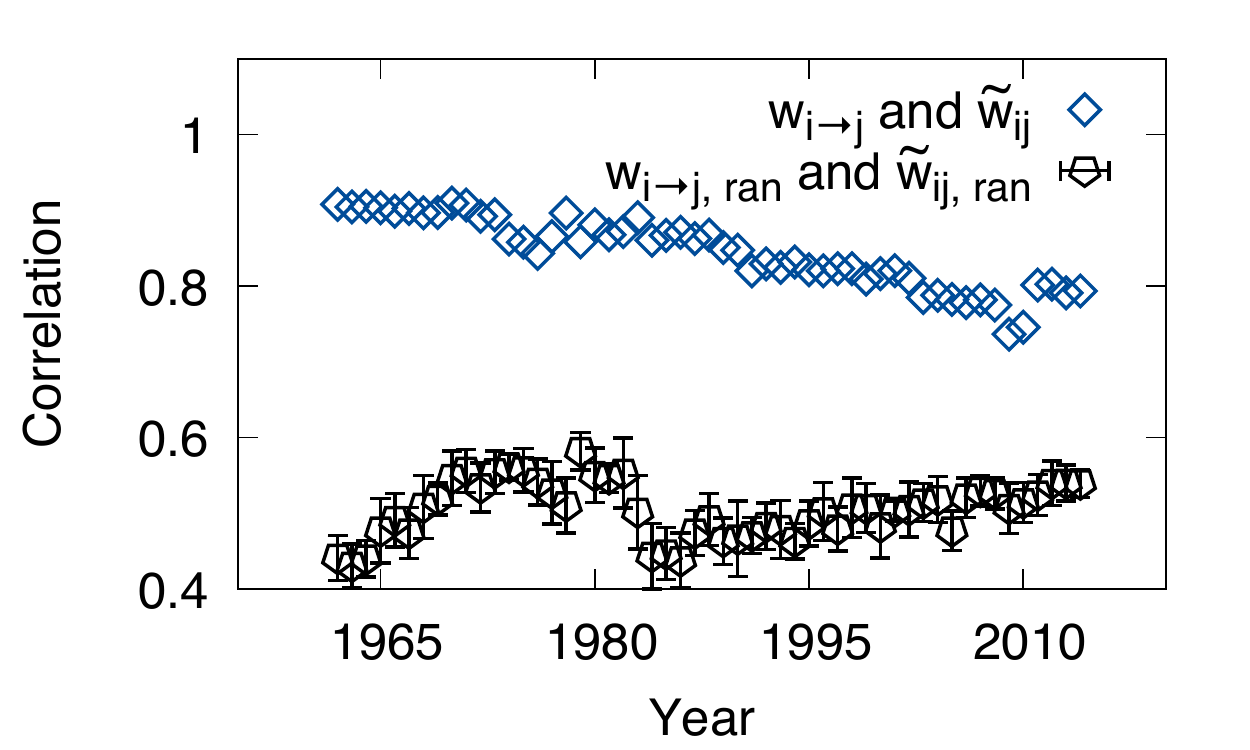}
\caption{The Pearson correlation coefficient between the normalized weight $\normw_{ij}$ and the original weight $w_{i\to j}$ in the WTW, for all of the existing directed edges in the subnetwork from the WTW. As a comparison, we also present the correlation coefficient for the randomized directed WTW networks, where the error bars indicate the standard deviation from $10$ randomized samples for each year.}
\label{fig:normwcorrel}
\end{figure}

Let us recall that the original WTW data is composed of \emph{directed} trade relations: imports and exports for bilateral trading nations, denoted by $w_{i\to j}\neq w_{j\to i}$ in general. So far, we have intentionally aggregated the weights as $w_{ij}\equiv w_{i\to j}+ w_{j\to i}$ regarded as a trade volume between two nations $i$ and $j$, as a test bed to extract directional information as described in Sec.~\ref{subsec:empricial_data}. In this section, we finally check if our method has successfully uncovered the genuine directional information by comparing the result to the original data. To recap, there exist the original amount of export from nation $i$ to nation $j$ denoted by $w_{i \to j}$ and the normalized weight from $i$ to $j$ denoted by $\normw_{ij}=w_{ij} / s(i)$ representing the inferred dependency of $i$ on $j$. We calculate the Pearson correlation coefficient between $w_{i \to j}$ and $\normw_{ij}$ when there is the directed edge from $i$ to $j$ in both the original directed network and the directed subnetwork extracted from our method, i.e., when $w_{i \to j} \neq 0$ and node $j$ belongs the top $\mdshnk(i)$ neighbors of $i$ in terms of weights. From Fig.~\ref{fig:normwcorrel}, we can see that the inferred weights in the extracted subnetwork and the real weights in the original directed network are highly correlated, which verifies the validity of our method in estimating the mutual dependency. 

The accuracy of this estimation is compared to the case of randomized directed weights $w_{i \to j,\mathrm{ran}}$ from the original WTW data, as shown in Fig.~\ref{fig:normwcorrel}, where the correlation coefficients represent the comparison between the randomized directed weights $w_{i \to j,\mathrm{ran}}$ and the normalized weights $\tilde{w}_{ij,\mathrm{ran}}$ from their own undirected networks by taking the same merging procedure $w_{ij,\mathrm{ran}} = w_{i \to j,\mathrm{ran}} + w_{j \to i,\mathrm{ran}}$. Note that our method regenerates the directional information (the correlation coefficient $> 0.4$) even in that randomized version to a degree due to the fact that $w_{ij,\mathrm{ran}}$ includes the original information $w_{i \to j,\mathrm{ran}}$. However, the correlation is much weaker than the original WTW networks, which implies the randomization process destroys the intrinsic crucial information that our method uses to recover the directionality. Therefore, it indicates both the effectiveness of our method and the amount of hidden information available.

\subsection{Mutuality}
\label{subsec:mutuality}

\begin{figure}
\includegraphics[width=0.9\columnwidth]{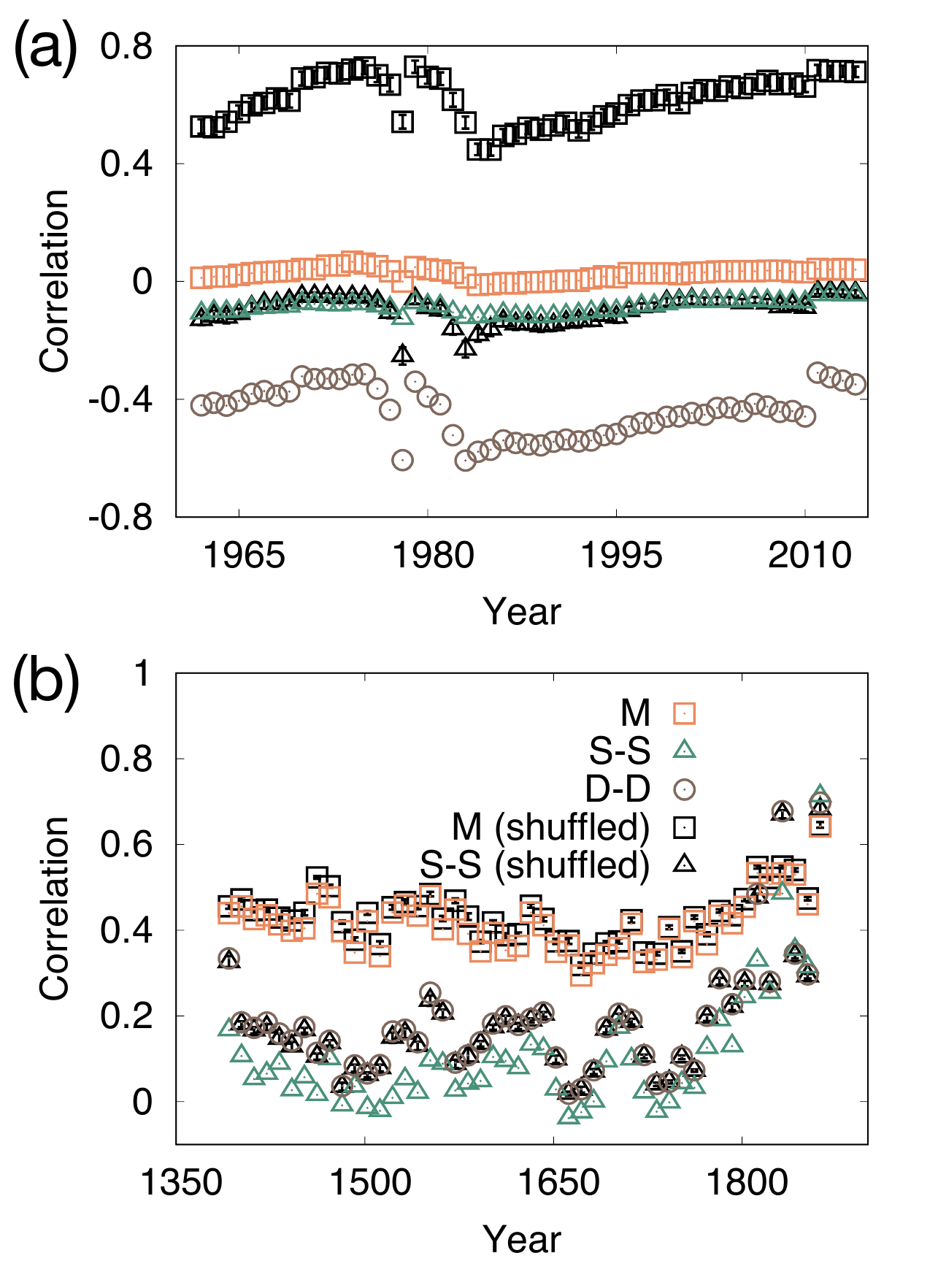}
\caption{Three types of correlations: the mutuality $M$, the degree-degree (D-D) correlation, and the strength-strength (S-S) correlation for (a) the WTW and (b) the AJD, and their null-model networks with shuffled weights, denoted by ``shuffled'' in the legends.}
\label{fig:corr}
\end{figure}

As the final analysis, we present the mutuality $M$ in Eq.~\eqref{eq:mutuality} and compare it with other pairwise correlation measures for structural properties. We have already argued that the mutuality can be subordinated to the underlying network structure in Sec.~\ref{sec:mutuality}---because the normalized weight is inversely proportional to the degree or the strength when the weights are homogeneous enough or random, the mutuality is expected to be correlated with the degree-degree (D-D) correlation~\cite{Assort2002,Assort2003} or the strength-strength (S-S) correlation. In Fig.~\ref{fig:corr}, we show the temporal changes of those correlation measures for the WTW and the AJD, along with those for their aforementioned null-model networks with randomized weights. 

First of all, in the case of WTW shown in Fig.~\ref{fig:corr}(a), one can check that the fluctuation of $M$ is much less severe than that of other correlations, in particular, compared to the large fluctuation of the D-D correlation in the late 70s to the early 80s when the substantial reorganization of international trade relations happened as discussed in Sec.~\ref{subsec:dependency}. In spite of the large structural changes reflected in the large fluctuation in the D-D correlation, the bilateral dependency reflected in $M$ has not been disrupted as severely as the network structure itself, so we speculate the situation as the following: in spite of turmoil in international trades caused by various geopolitical reasons, nations might have tried their best to quickly mitigate the shock and maintain the overall mutual dependence in response.

The implication of mutuality values themselves becomes clear when we compare them with the results from the null-model networks. Again, we generate 100 null-model networks with completely shuffled weights on the original network structure, calculate the mutuality and the S-S correlation (the D-D correlation would be the same because the network structure itself is not altered), and plot their ensemble-averaged values in Fig.~\ref{fig:corr} in addition to the correlation values from the original networks. The most prominent difference between the original network and its null model is observed in the case of mutuality of the WTW in Fig.~\ref{fig:corr}(a), and in particular, the mutuality of the WTW is much smaller than that of its null model. This again confirms our previous conclusion that the international trade is less mutual, as discussed in Sec.~\ref{subsec:dependency} and Fig.~\ref{fig:reciprocity}(b). 
The positive values of $M$ in the case of null models have the same origin as the larger reciprocity discussed in Sec.~\ref{subsec:result_effk}. The average and variance of local link weights are not distinguishable between the two end nodes of a link in the null models. Thus a link with high (low) weight is likely to have similar normalized weights commonly larger (smaller) than the average $\mu$. It does not hold for the real WTW, in which the scales of the link weights of two connected nodes may be quite different, and thus the normalized weight of a link from the viewpoint of one end node may be much different from the other, reducing mutuality. For more discussions and examples, see Appendix~\ref{sec:app_compare_m_r}.

The absence of significant effects of weights and the results from it in the AJD is reconfirmed with the mutuality and other correlations as well, as shown in Fig.~\ref{fig:corr}(b). As expected, the mutuality of the AJD is quite similar to that of the corresponding null-model networks, not surprisingly because of their relatively uniform weights, the details of which are already discussed in Sec.~\ref{subsec:dependency}.

\section{Summary and Discussion}
\label{sec:discussion}

We have proposed the framework for constructing a directed subnetwork composed of the most essential edges via the concept of information entropy, based on heterogeneity of local distributions of weight around each node.
We call the number of such essential neighbors of a node the effective out-degree, which plays the role of a local or ego-centric threshold of extracting the most important neighbors \emph{from} the node. This naturally appearing but initially hidden directionality from each of individual nodes is the cornerstone of our framework. Although we have focused on the case of the Shannon entropy ($\alpha \to 1$) almost exclusively in our work, by tuning the parameter $\alpha$ one can control the overall sensitivity of the threshold. 
To demonstrate the utility of our method, we have compared two series of real networks composed of temporal snapshots: the WTW and the AJD, followed by the comparison with their weight-randomized version as the null model. We have analyzed the hidden dependency within the networks by taking both the global- and the local-scale properties and concluded that the WTW has intrinsically less mutual or unequal bilateral dependency between the nations, while people in the AJD are connected with more mutual dependency from their narrowly distributed weights. In addition, we have verified that our method extracts the most essential directed relation by comparing the result with the original directional information (export and import) in the WTW.

We can apply the extracted directed subnetwork to various purposes, depending on the context. In general, the directionality from $i$ to $j$ in our framework indicates the dependency of $i$ on $j$, so it effectively captures the flow from less influential nodes to more influential nodes, roughly speaking. In social relations, for instance, the directionality may insinuate the hidden authoritative relations among nominally mutual ``friendship.'' Another example is various types of biochemical networks, where seemingly ``related'' chemical/metabolic reactions or genetic entities could in fact hide their true identity of asymmetric dependency, which would enable us to prioritize a part of networks to engineer the system better, e.g., when we try to find a new drug target. Beyond the inference to the directionality in static networks, we may utilize the fact that the directional information connotes the temporal information as any type of interaction takes time. Therefore, albeit not perfectly, the directionality may help us to deduce the temporal order from temporally accumulated networks as well, which would be of great importance when it comes to reconstruction of causality or the Bayesian formulation~\cite{causality}. In the viewpoint of dynamical processes on networks, the cascading effect from concatenation of such a directionality may provide a crucial hint to infer the long-range effective flow in networks~\cite{Ghavasieh2020}.

Finally, to take a more concrete example, our analysis of the WTW has demonstrated the potential of our method to applications to economic and other sectors dealing with global problems as well, we believe. We would like to emphasize that pointing out specific nations with characteristic properties in terms of LR and AR is much more meaningful than just providing interesting anecdotal examples, because each of the interrelationships in the WTW actually affects our daily life. The hidden directionality in epidemic spreading processes can be crucial to detect superspreaders or superblockers, which is tightly related to the global economy as now all of us know. For example, one can measure conventional centrality measures such as the closeness or perform community detection or $k$-core decomposition in the directed subnetwork from an original network, compared with those in the original network. In the case of closeness centrality, the average closeness from a node to other nodes in the directed subnetwork indicates the node's effective long-range proximity by considering the most relevant paths. We hope to sharpen our tool more to prepare for more practical applications on top of a more solid theoretical background in the future.

\begin{acknowledgments}
The authors thank Pan-Jun Kim (김판준) for initiating the idea of using entropy-based measures to quantify the heterogeneity in the weight distribution of networks, during the collaboration~\cite{SHLee2010}. The National Research Foundation (NRF) of Korea grant funded by the Korean Government supported this work through Grant Nos.~NRF-2021R1C1C1007918
 (M.J.L.), NRF-2017R1A2B3006930 (B.L. and H.J.), NRF-2019R1A2C1003486 (D.-S.L.), NRF-2018R1C1B5083863 (S.H.L.), and NRF-2021R1C1C1004132 (S.H.L.).
\end{acknowledgments}

\appendix

\section{More detailed description of the empirical data}
\label{sec:app_description}

\begin{figure*}
\includegraphics[width=0.8\textwidth]{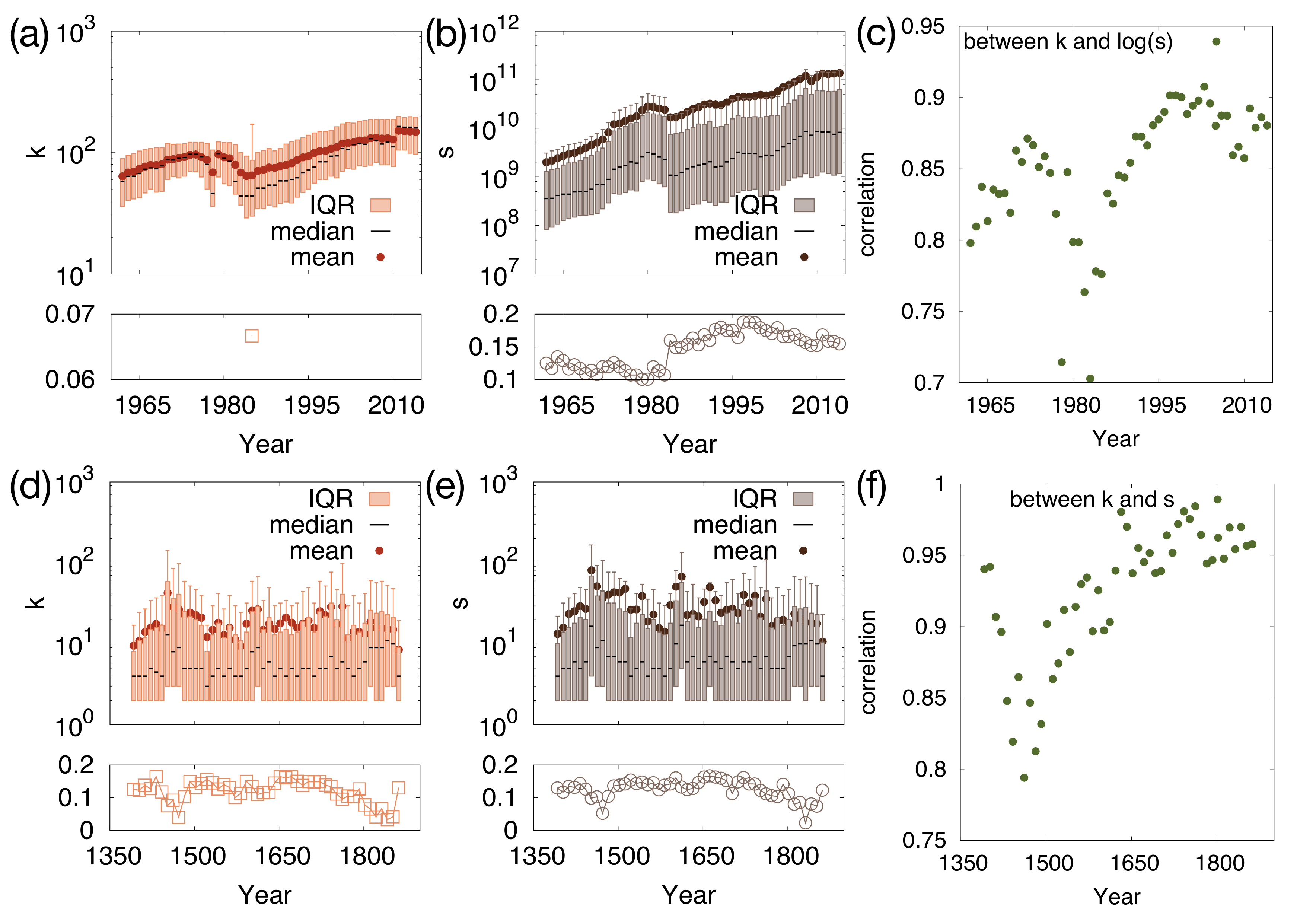}
\caption{The temporal changes of the degree and strength distributions of the WTW (top) and the AJD (bottom), with the same format as in Fig.~\ref{fig:empiricaldata}. The distributions of [(a) and (d)] the degree $k$, [(b) and (e)] the strength $s$ are shown, and we also present [(c) and (f)] the Pearson correlation coefficient between them.}
\label{fig:app_boxplot}
\end{figure*}

\subsection{The world trade web}
The world trade web (WTW)~\cite{Hidalgo2007,Hidalgo2009,Hidalgo2011AAAI,WTWdata} is historically recorded for 53 years from 1962 to 2014, and we use the annually aggregated networks (so 53 networks in total) in our analysis. The nations participating in the international trade are the nodes, and the trade relations are the edges with the weights corresponding to trade volumes. There were 152 nations in 1962, and the number of nations had increased to 233 in 2014. The trade data contain the import and export amount of the products in the unit of United States (US) dollars. The exported or imported products are classified by the international standard. More specifically, for the data from 1962 to 2000, the trades are classified with Standard International Trade Classification (SITC). For more recent data (2001--2014), the data is provided by United Nations (UN) Comtrade Database.
As we have explained in Sec.~\ref{subsec:empricial_data}, we first merge the export and import sides and treat them as undirected weighted edges to test our method and then compare the result with the real directional trade information in Sec.~\ref{subsec:inference_to_original_network}.

\subsection{The Annals of the Joseon Dynasty}
The Annals of the Joseon Dynasty (AJD) is a historical record written in classical Chinese, ordered chronologically. It covers the 472 years (1392--1863) corresponding to the reigns of 25 kings. The AJD provides plentiful information about not only political activities at the royal court, but also economic, social, and cultural events of the Joseon Dynasty. The National Institute of Korean History runs a web service that offers both the original Chinese text and its translated version in Korean.
The structure of the AJD is as follows: each reign is composed of the record of years, the record of each year comprises the record of months, the record of each month comprises the record of days, and the record of each day contains articles. The data consists of 6992 months, 143\,066 days, and 380\,009 articles. The entire data set was extracted from the official website~\cite{AJD,AJDdata}. In the AJD, there are 54\,526 number of people manually tagged by modern historians, and they are the nodes in the network. As described in Sec.~\ref{subsec:empricial_data}, the edges between node pairs represent the number of sentences mentioning the pair together within a ten-year time window.

\subsection{Basic local properties of the networks}
As mentioned in Sec.~\ref{subsec:empricial_data}, we add the distributions of the basic local network properties: the degree $k$ and the strength $s$, followed by the correlation between them. In Fig.~\ref{fig:app_boxplot}, we show the temporal changes of $k$ and $s$ in the same format (the mean, the median, the IQR, and the outliers) as in Fig.~\ref{fig:empiricaldata}, and their correlation. 
The degree distributions of the WTW shown in Fig.~\ref{fig:app_boxplot}(a) are relatively homogeneous characterized by their well-defined representative mean values, judged by their similarity to the median and the (almost) absence of outlier. Except for that, all of the other distributions (the strength distributions of the WTW and the degree and strength distributions of the AJD) are quite heterogeneous; they are severely right-skewed with non-negligible outliers, as shown in Figs.~\ref{fig:app_boxplot}(b), \ref{fig:app_boxplot}(c), and \ref{fig:app_boxplot}(d). 
They are right-skewed distributions inferred by mean values larger than medians with the large fraction of outliers. Overall, the temporal fluctuations of the degree and the strength resemble each other for a given data, which is also consistent with the large correlation coefficients between them in Figs.~\ref{fig:app_boxplot}(c) and \ref{fig:app_boxplot}(f). Note that we calculate the correlation between $k$ and $\log (s)$, because the correlation is larger than that between $k$ and $s$ itself, which means the strength exponentially (or at least superlinearly) increases with the degree $k$, roughly speaking. In the AJD, the correlation between $k$ and $s$ is close to linear, which is also reflected in the similarity between Figs.~\ref{fig:app_boxplot}(d) and \ref{fig:app_boxplot}(e).

\section{Comparison with the multiscale backbone method}\label{sec:MBM}

\begin{figure*}
\includegraphics[width=\textwidth]{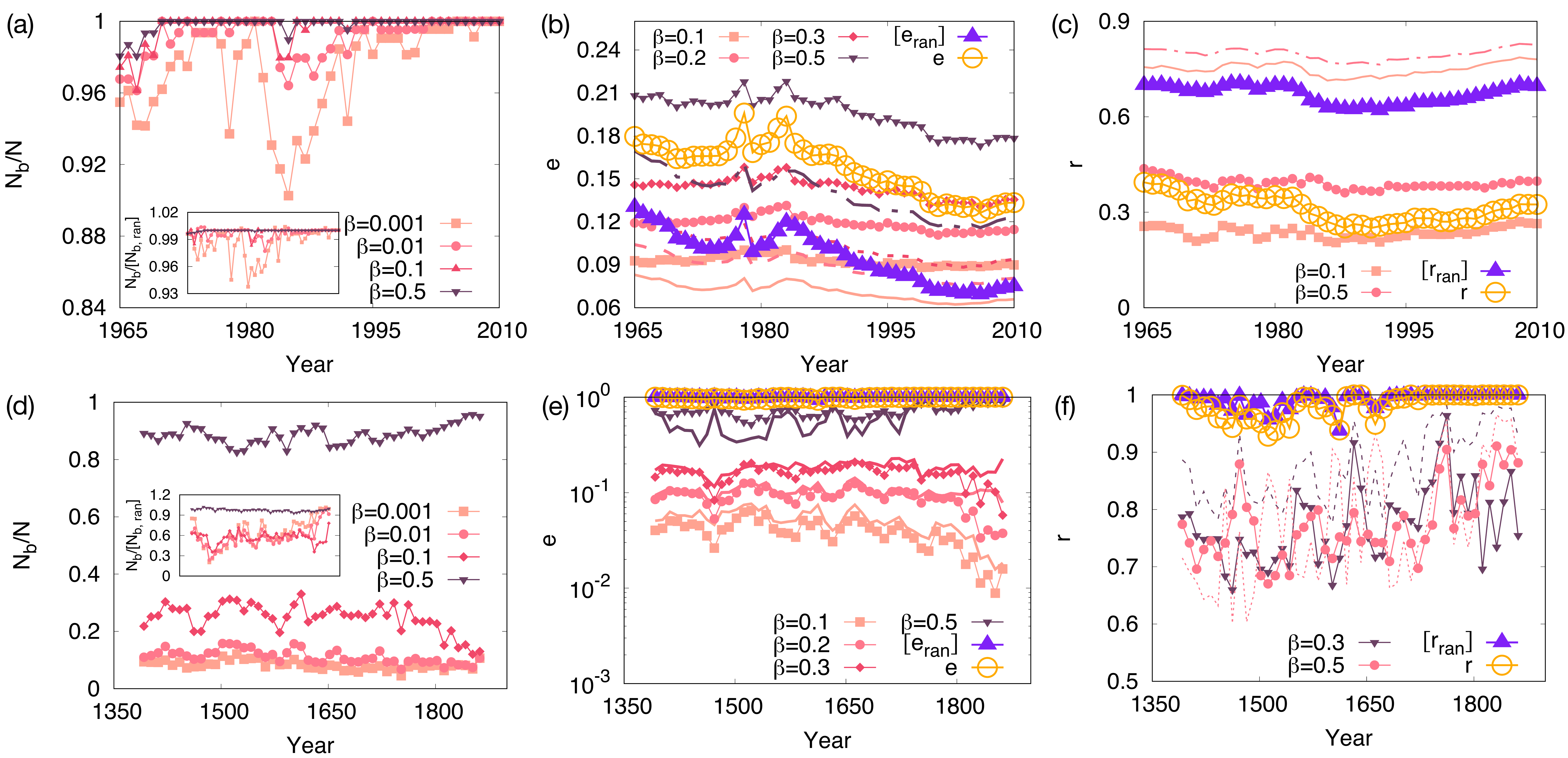}
\caption{Global measures for the directed subnetwork by the MBM~\cite{Serrano2009}, compared with our method, for [(a)--(c)] the WTW and for [(d)--(f)] the AJD. (a) and (d) show the survival fraction $N_b/N$ of nodes from the MBM with $\beta=0.001, 0.01, 0.1$, and $0.5$. In our method, $N_b/N=1$ always by definition. The ratio of the number $N_b$ of survival nodes in the original network to the average number $[N_{b, {\rm ran}}]$ of the survival nodes in the weight-shuffled networks is plotted in the insets. (b) and (e) show the relative edge density $e$, and (c) and (f) show the global reciprocity $r$. The results from our method represented by the orange open circles and purple filled squares are taken from Fig.~\ref{fig:reciprocity}. The smaller filled symbols represent the results from the MBM with given threshold values $\beta=0.1, 0.2, 0.3, $ or $0.5$. For each case, the results from the weight-shuffled null-model networks for each given value of $\beta$ are shown with the solid or dashed curves characterized by the same color taken from the same $\beta$ value for the real data.}
\label{fig:app_serrano}
\end{figure*}

In this Appendix, we compare our method to another extraction method. We have mentioned the methods such as the minimum (maximum) spanning tree, global threshold, and the multiscale backbone method (MBM)~\cite{Serrano2009} in the main text. The most important aspect in our extraction method is to excavate the hidden directionality from originally undirected networks. As the spanning tree and global thresholding are in principle unable to do that, we focus on the MBM for comparison. In fact, we would like to emphasize that the original paper introducing the MBM~\cite{Serrano2009} did \emph{not extract the directional information}, by accepting statistically significant edges from \emph{either} direction; therefore, we modify the original MBM to allow extraction of directional information in an intuitive way for fair comparison.

The MBM~\cite{Serrano2009} first utilizes the normalized weight $\normw_{ij}$ in Eq.~\eqref{eq:wtilde} as well. Then, for an edge between nodes $i$ and $j$, one computes the probability $\beta_{ij}$\footnote{The original paper~\cite{Serrano2009} uses the symbol $\alpha$, but we modified it to $\beta$ to avoid the possible confusion with the parameter $\alpha$ used in this paper for the R{\'e}nyi entropy.} for the normalized weight from node $i$ to node $j$ under the assumption that randomly distributed weights are larger than the original weight $w_{ij}$. If $\beta_{ij} < \beta$ or $\beta_{ji}<\beta$ with $\beta$ being the significance level, one considers that the edge is significant and keeps it as a component of the network backbone. Otherwise, the edge is removed at the backbone. The value of $\beta$ here controls the level of stringency as small values of $\beta$ indicate stricter criteria for an edge to survive in the backbone. 
We can simply modify the MBM by treating $\beta_{ij}<\beta$ and $\beta_{ji}<\beta$ as \emph{separate} conditions to extract the directed backbone as our method. Before we present the results using this modified version for the network data we presented in the main text, we remark on an important difference of the MBM compared with our method to highlight their contrasting viewpoints. First, as we already discussed in the last of Sec.~\ref{subsec:effective_outdegree}, for the uniform weight distribution $\normw_{ij}=1 / k(i)$, all edges connected to node $i$ are deemed as \emph{equally important edges} in our method (so all of them are retrieved) but they are \emph{equally insignificant edges} in the MBM (so all of them are removed). In such a case, the node with the locally uniform weight distribution is isolated one in the backbone in the MBM and removed accordingly. This can happen even for nodes with non-uniform local weight distributions in the MBM, depending on $\beta$.

\subsection{Global measures for the WTW and the AJD}
First, we present representative global measures of the directed subnetwork using the aforementioned modified version of MBM with various threshold values applied to the WTW and AJD data in Fig.~\ref{fig:app_serrano}, along with those measures from our method (taken from Fig.~\ref{fig:reciprocity}, in other words), in comparison to the corresponding weight-shuffled null models. The upper and lower panels correspond to WTW and AJD, respectively. As the MBM removes a fraction of nodes once they are isolated as described before, we first plot the number $N_b$ of remaining node in the directed subnetwork as their fraction with respect to the number $N$ of original nodes [Figs.~\ref{fig:app_serrano}(a) and \ref{fig:app_serrano}(d)].
Most nodes are retrieved in the directed subnetwork of WTW, as long as the threshold $\beta$ for the MBM is not too small. In sharp contrast, large fractions of nodes are discarded in the directed subnetwork of AJD, which dramatically demonstrates the difference between our method and the MBM---the MBM tends to filter out the edges with similar $\normw_{ij}$ because they are \emph{equally unimportant} as previously discussed. A similar level of node survival to the WTW is achieved for $\beta \gtrsim 0.5$ in the case of AJD, which will correspond to rather too generous a criterion as the significance level.
The ratio of the surviving nodes in the original network to that for the $100$ weight-shuffled null-model networks is plotted in the insets of Figs.~\ref{fig:app_serrano}(a) and~\ref{fig:app_serrano}(d). Based on the observation that $N_b$ for the AJD is notably smaller than the value for their randomized counterparts when $\beta < 0.5$, we can see that the small survival rate of the nodes in the AJD is caused by their intrinsic local correlation between the normalized weights.

The relative edge density $e$ and the global reciprocity $r$ for the directed subnetworks from the MBM with various values of the confidence level $\beta$ [Figs.~\ref{fig:app_serrano}(b), \ref{fig:app_serrano}(c), \ref{fig:app_serrano}(e), and \ref{fig:app_serrano}(f)] show similar behaviors to those from our method presented in Fig.~\ref{fig:reciprocity}. For direct comparison, the results from our method are also plotted (with the same symbols and colors as in Fig.~\ref{fig:reciprocity}) in addition to the MBM results there. As expected, the relative edge density $e$ calculated for the MBM-based directed subnetwork becomes smaller as more strict criteria for selecting edges (larger $\beta$ values) are applied [Figs.~\ref{fig:app_serrano}(b) and~\ref{fig:app_serrano}(e)] for both data, but the amount of decrement is much larger for the AJD. The trend is not surprisingly understandable from the same argument presented in the case of $N_b / N$---the MBM filters out ``equally unimportant'' edges. In comparison, the directed subnetwork from our method for the WTW has roughly similar values of $e$ to the ones from the MBM with $0.3 \leq \beta \leq 0.5$ in the early period and later to the ones with $\beta = 0.3$ as shown in Fig.~\ref{fig:app_serrano}(b). The relative edge density for the directed subnetwork of the AJD with the MBM clearly suffers from the significantly reduced number of nodes as shown in Fig.~\ref{fig:app_serrano}(e). 
Despite the sharp contrast in treating edges with locally homogeneous weights, our method and the MBM share a similar behavior of $e$ when it comes to the comparison with the weight-shuffled null models. As in our method in Fig.~\ref{fig:reciprocity} [and Figs.~\ref{fig:app_serrano}(b) and \ref{fig:app_serrano}(e) again], the average $e$ values from $100$ weight-shuffled null-model networks for the MBM are also depicted for each value of $\beta$, represented by a solid or dashed curves with the same color used to plot $e$ values for the real data with the same $\beta$ value. Similar to the result from our method, in the case of MBM as well, the $e$ values are systematically larger for the real data than the weight-shuffled null models for the WTW, and the real data and null models are similar for the AJD. In other words, the fact that the WTW harbors more intricate directional correlations than the AJD (discussed in Sec.~\ref{subsec:dependency}) is cross-checked with the MBM. 

In Figs.~\ref{fig:app_serrano}(c) and~\ref{fig:app_serrano}(f), the global reciprocity $r$ with $\beta=0.1$ and $\beta=0.5$ for the WTW and $\beta=0.3$ and $\beta=0.5$ for the AJD is plotted in the same way as the relative edge density $e$. In the WTW, the value of $r$ for $\beta=0.1$ is lower than that for $\beta=0.5$ [Fig.~\ref{fig:app_serrano}(c)], and the reciprocity from our method across different years lies between the two cases of the MBM.  
Note that the reciprocity values for the real WTW data are significantly lower than their weight-shuffled null models, both in our method and the MBM. In contrast, the reciprocity values are similar for the real AJD data and their weight-shuffled null models both in our method and the MBM. Therefore, this again cross-checks the fact that the WTW is composed of intrinsically unequal relationships between trading nations, as discussed in Sec.~\ref{subsec:dependency}.

\subsection{Local properties of the directed subnetwork of the WTW}

\begin{figure*}
\includegraphics[width=\textwidth]{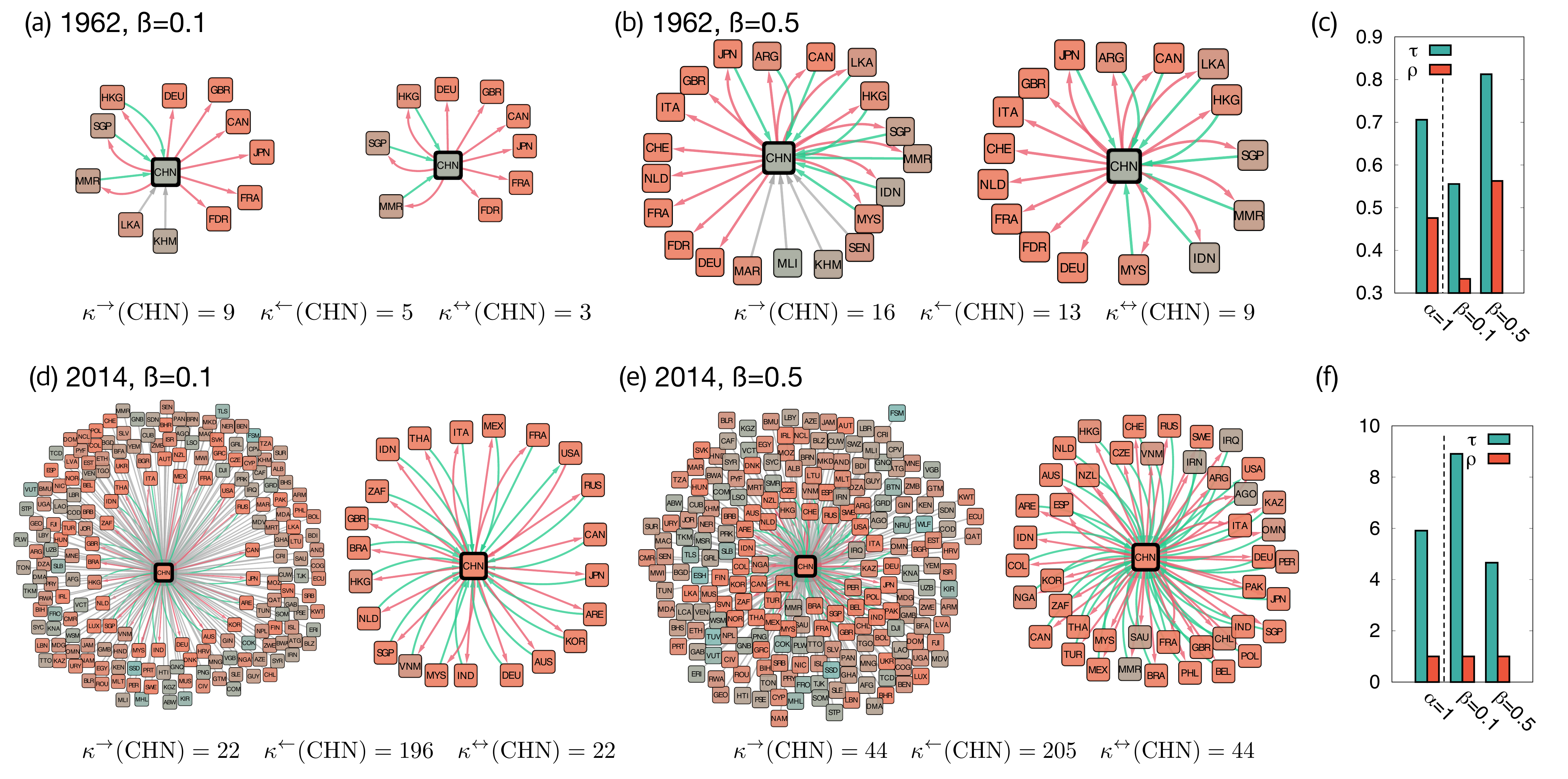}
\caption{Parts of subnetworks of the WTW in the CHN-centric viewpoint obtained from the MBM in [(a) and (b)] 1962 and [(d) and (e)] 2014 with the confidence level [(a) and (d)] $\beta=0.1$ and [(b) and (e)] $\beta=0.5$. The colors of nodes and edges are selected in the same way as in Fig.~\ref{fig:network}. The changes of the AR $\tau$ and the LR $\rho$ are shown in [(c) and (f)]. The case of $\alpha=1$ corresponds to our method, and $\beta=0.1$ and $\beta=0.5$ represent the MBM with those values of $\beta$.}
\label{fig:app_network_china}
\end{figure*}

\begin{figure*}
\includegraphics[width=\textwidth]{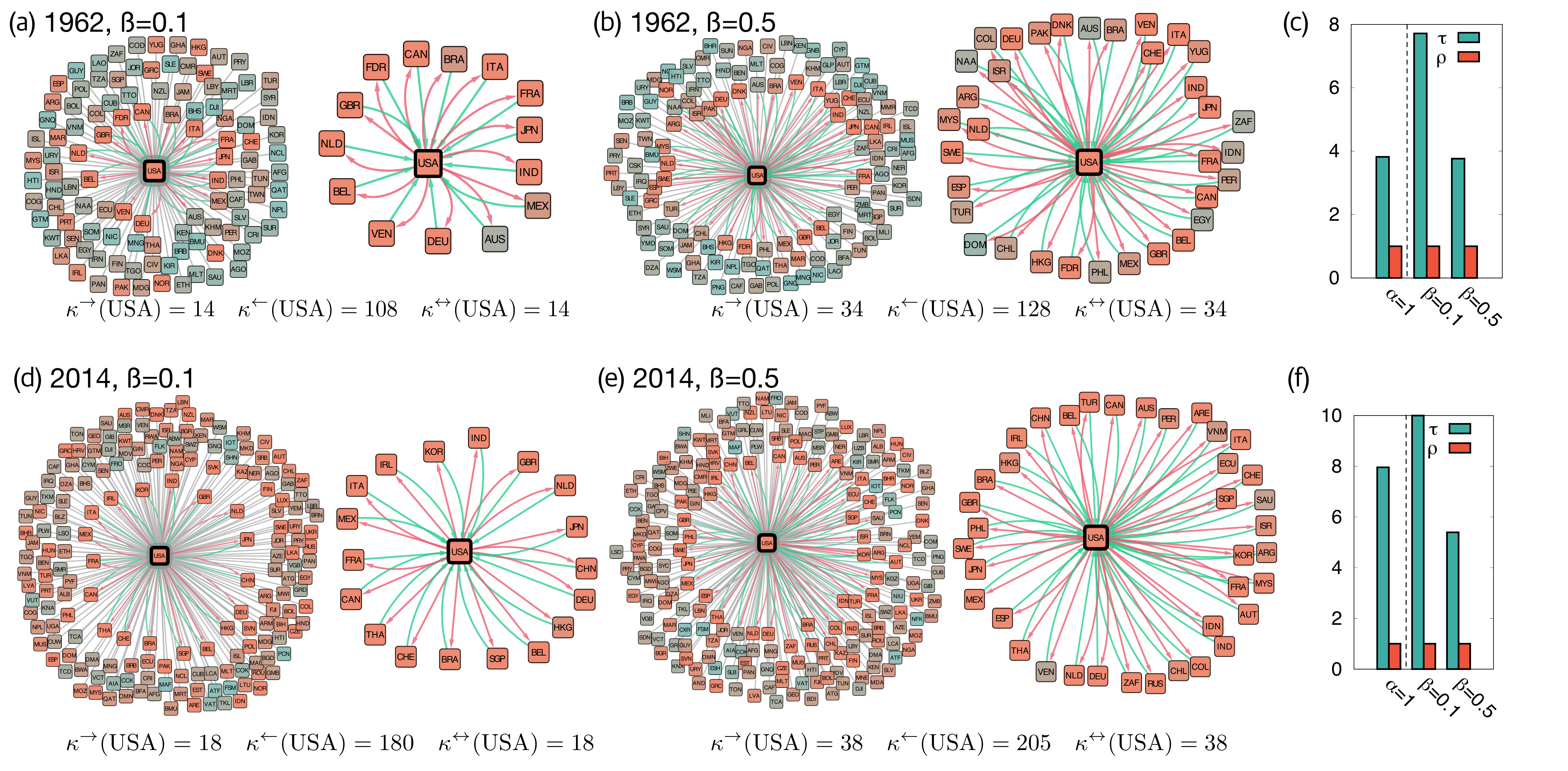}
\caption{Parts of subnetworks of the WTW in the USA-centric viewpoint obtained from the MBM in [(a) and (b)] 1962 and [(d) and (e)] 2014 with the confidence level [(a) and (d)] $\beta=0.1$ and [(b) and (e)] $\beta=0.5$. See the caption of Fig.~\ref{fig:app_network_china} for details.}
\label{fig:app_network_usa}
\end{figure*}

In the main text, we have explored the local properties of the directed subnetwork of the WTW generated from our method. We take the MBM to perform the same type of analysis for the WTW as comparison. As depicted in Fig.~\ref{fig:local_dependency}, we choose the years 1962 and 2014 and present the results for the CHN-centric and USA-centric subnetworks using different values of confidence level $\beta=0.1$ and $\beta=0.5$ for the MBM in Figs.~\ref{fig:app_network_china} and \ref{fig:app_network_usa}. As expected, smaller numbers of edges are kept at a more strict criterion ($\beta=0.1$) for a significant link than at a less strict criterion ($\beta=0.5$).

In Figs.~\ref{fig:app_network_china}(c), \ref{fig:app_network_china}(f), \ref{fig:app_network_usa}(c), and \ref{fig:app_network_usa}(f), the local properties: the AR $\tau$ and the LR $\rho$ values of CHN and USA are compared with those in the main text. The label $\alpha=1$ stands for the case of our method using the Shannon entropy, and the labels $\beta=0.1$ and $\beta=0.5$ indicate the case of the MBM with those values of confidence level. 
Notably, $\tau < 1$ for CHN in 1962 from both our method and the MBM with two different $\beta$ values, and $\tau > 1$ for CHN in 2014 and for USA in both 1962 and 2004.
Accordingly, it is consistently confirmed by both methods that the CHN becomes an overwhelming trading partner than the past over a few decades and that the USA has always been an important trading partner in the world trade.

\begin{figure}
\includegraphics[width=\columnwidth]{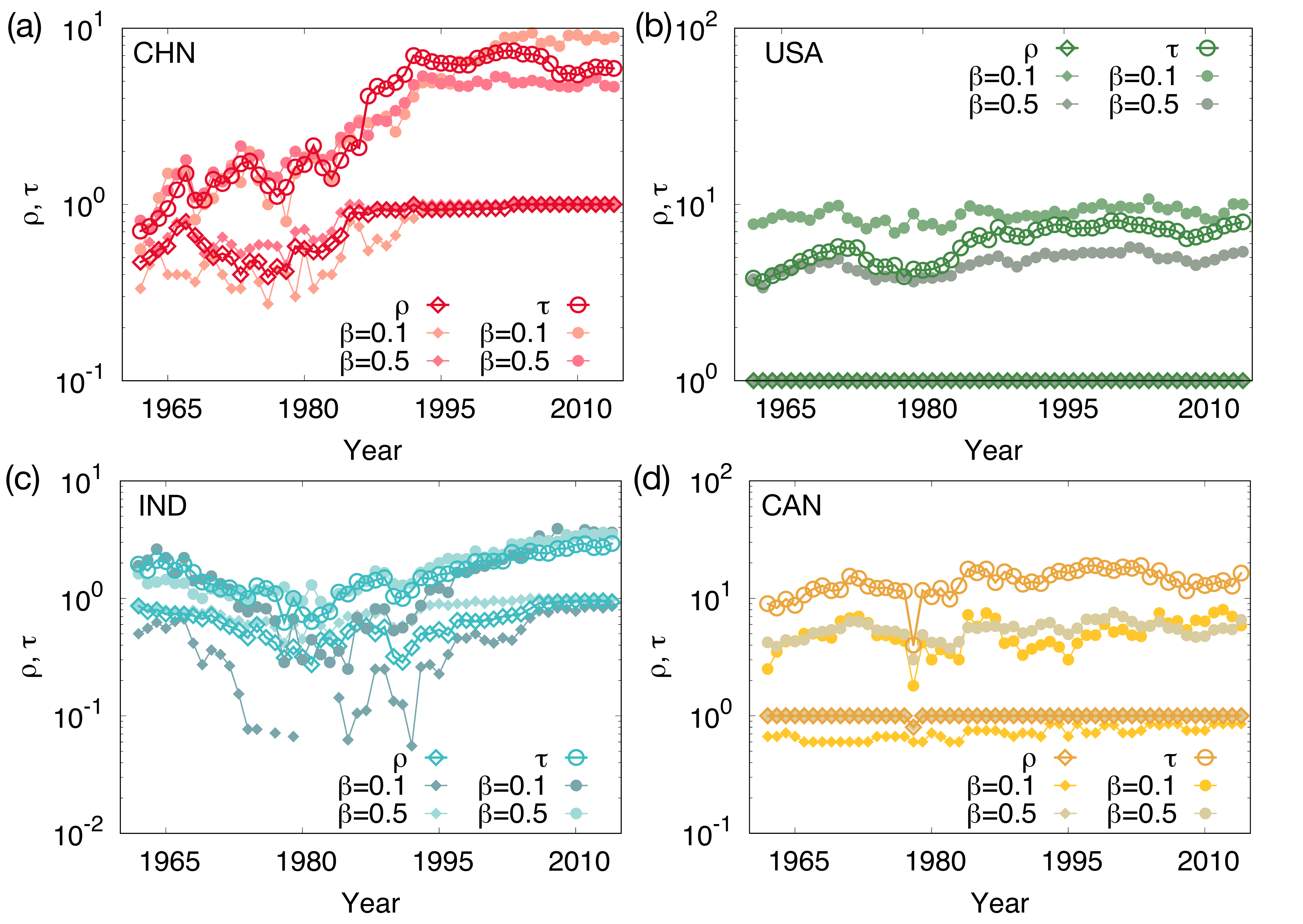}
\caption{The time series of the LR $\rho$ and AR $\tau$ values for selected countries: (a) CHN, (b) USA, (c) IND, and (d) CAN. The open and filled symbols indicate the result obtained from our method and the MBM, respectively. The diamond and circle symbols indicate $\rho$ and $\tau$, respectively. Note that we only show the point with $\mdshnk\neq 0$.}
\label{fig:app_wtw_local}
\end{figure}

As the final analysis with respect to the local property, we show the time series of $\rho$ and $\tau$ values for four selected nations [as shown in Figs.~\ref{fig:local_dependency}(d) and \ref{fig:local_dependency}(e)] in Fig.~\ref{fig:app_wtw_local}. The results from our method as in Fig.~\ref{fig:local_dependency} are replotted for direct comparison. The overall temporal trends are similar for the results from the MBM with different $\beta$ values. The $\tau$ values are almost always close to unity for USA and CAN, and they are approaching unity as time goes by for CHN and IND. For the case when $\tau \gtrsim 1$, the $\rho$ values are close to $1$. IND has a larger gap between $\rho$ and $\tau$ than the cases of the other nations, and for $\beta = 0.1$ there are years when $\rho = 0$ and $\tau > 1$ for IND. It represents the non-existence of reciprocal important trading partners among its incoming friends, which corresponds to totally asymmetric trading relationships. Such a striking case is not observable from our method or the MBM with large $\beta$ values, so it demonstrates the importance of trying different methods to analyze this type of data.

\section{Understanding the null-model networks via synthetic networks}\label{sec:app_synthetic}

\begin{figure}
\includegraphics[width=\columnwidth]{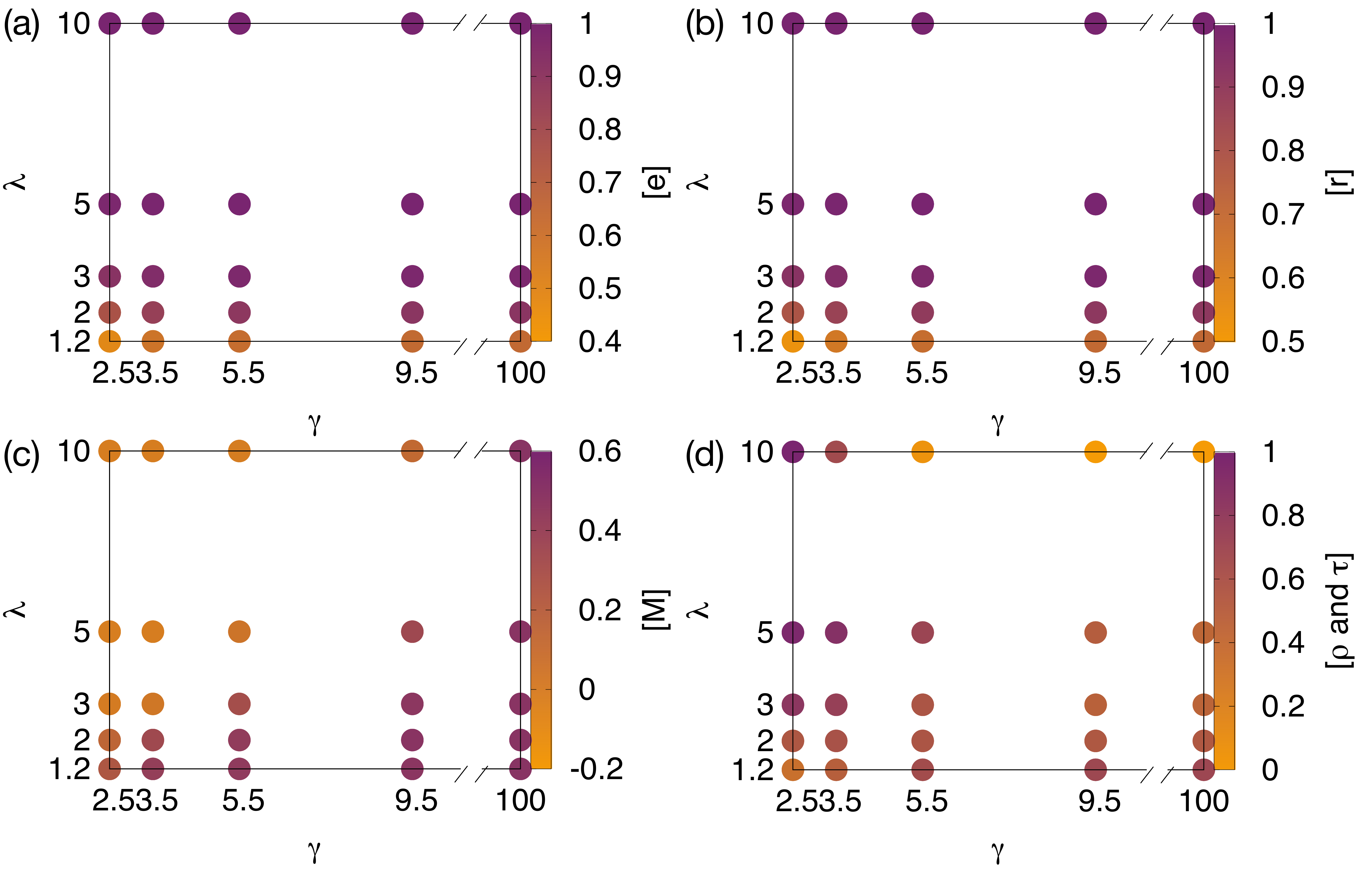}
\caption{The results for $100$ model networks with given degree and weight distributions in Eqs.~\eqref{eq:app_pk} and \eqref{eq:app_pw}, respectively. The shade of each point indicates the values of (a) the average relative edge density $e$, (b) the global reciprocity $r$, (c) the mutuality $M$,  and (d) the correlation between the LR $\rho$ and the AR $\tau$.}
\label{fig:app_synthetic}
\end{figure}

In the main text, we have used the weight-shuffled version of empirical network data as the null-model counterpart. The weight-shuffled networks are effective at comparing the original networks' intrinsic correlation between weights with the case of completely destroyed version, with the (binary) connection structure intact. There can be different levels of a null model, of course, due to a number of different network parameters. In this Appendix, we take perhaps the most elementary version of null-model networks: model networks only characterized by independently distributed degrees on nodes and weights on edges, to deduce the most basic property of such distributions in the scheme of our analysis.

For the model networks, the only assumption is based on the fact that many real-world networks are arguably\footnote{We invite interested readers to check Refs.~\cite{Broido2019,Voitalov2019} and judge the situation themselves.} close to scale-free with heterogeneous weight distributions, with the power-law form of distribution, i.e.,
\begin{align}
    P_d(k) &\sim k^{-\gamma}  \label{eq:app_pk}\\
    P_w(w) &\sim w^{-\lambda} \label{eq:app_pw},
\end{align}
where $k$ and $w$ are the degree and the weight, respectively. Thus the degree and weight exponents, $\gamma$ and $\lambda$, respectively, adjust the heterogeneity of the corresponding quantities. We first construct our model scale-free networks using the configuration model~\cite{configmodel} with the number of nodes $N=10\,000$ and the fixed mean degree $\langle k \rangle = 2$, for various values of the exponents $\gamma=2.5, 3.5, 5.5, 9.5$ (for different levels of heterogeneity), and $100.0$ (for the extreme case of homogeneity). On top of such substrate structures, we assign weights on each edge independently from the power-law distribution in Eq.~\eqref{eq:app_pw} with $\lambda=1.2, 2.0, 3.0, 5.0$, and $10.0$. For different combinations of these exponents, we take the same procedure of calculating the normalized weights in Eq.~\eqref{eq:wtilde} and extract the directed subnetworks with the effective out-degree in Eq.~\eqref{eq:effective_degree_shannon}. From the procedure and resultant directed subnetworks, we show the measures of our main interest: the global relative edge density $e$ in Eq.~\eqref{eq:edgedensity}, the global reciprocity $r$ in Eq.~\eqref{eq:reciprocity}, the mutuality $M$ in Eq.~\eqref{eq:mutuality}, and the correlation between local reciprocity $\rho$ in Eq.~\eqref{eq:rho} and the attraction ratio $\tau$ in Eq.~\eqref{eq:tau} for the individual nodes, introduced in the main text\footnote{They correspond to Figs.~\ref{fig:reciprocity}, \ref{fig:local_dependency}(c), and \ref{fig:corr} for the empirical data and their weight-shuffled versions.} in Fig.~\ref{fig:app_synthetic}.

The relative edge density $[e]$ in Eq.~\eqref{eq:edgedensity} and the global reciprocity $[r]$ in Eq.~\eqref{eq:reciprocity}, averaged over $100$ networks generated for each of the aforementioned $(\gamma,\lambda)$ combinations, become notably larger for larger values of $\lambda$ (more homogeneous weights) up to the point where the values approach unity as shown in Figs.~\ref{fig:app_synthetic}(a) and~\ref{fig:app_synthetic}(b).  The result is straightforward to understand as the overall homogeneity in weights just increases the chance for each of the $(i,j)$ and $(j,i)$ pairs for the original edges $(i,j)$ to be taken in the directed subnetwork according to our entropy-based effective out-degree in Eq.~\eqref{eq:effective_degree}. The heterogeneity in degree, by contrast, cannot play any role here because they just control the topological position of each edge.

The mutuality $M$ in Eq.~\eqref{eq:mutuality} and the correlation between the LR values $\rho$ and the AR values $\tau$ in Eqs.~\eqref{eq:rho} and \eqref{eq:tau}, however, are affected by both $\gamma$ and $\lambda$ values, as shown in Figs.~\ref{fig:app_synthetic}(c) and \ref{fig:app_synthetic}(d).
As described in the last paragraph of Sec.~\ref{sec:mutuality}, the mutuality strongly depends on the underlying network structure (assortativity, for instance) and of course on the weight distribution as in the case of the relative edge density and the global reciprocity. The overall trend of mutuality shown in Fig.~\ref{fig:app_synthetic}(c) indicates that more homogeneous degree distributions and more heterogeneous weight distributions (the lower right part of the plot) induce larger values of mutuality. The effect of weight distribution is understandable by first noting that the pair of normalized weights $\tilde{w}_{ij}$ and $\tilde{w}_{ji}$ stem from the same original weight $w_{ij}$, so without the nontrivial correlation between $s(i)$ and $s(j)$ it is likely that $\tilde{w}_{ij}$ and $\tilde{w}_{ji}$ are positively correlated in general. The level of the correlation is then determined by the overall variance in the weight distribution, and larger ``background'' variance of the weights (corresponding to small values of $\lambda$) makes the baseline correlation between $\tilde{w}_{ij}$ and $\tilde{w}_{ji}$ more prominent. The heterogeneity in degree (smaller $\gamma$ values) seems to weaken the baseline correlation between $\tilde{w}_{ij} = w_{ij} / s(i)$ and $\tilde{w}_{ji} = w_{ij} / s(j) = \tilde{w}_{ij} s(i)/s(j)$ by imposing heterogeneous values of $s(i)$ and $s(j)$. In the most extreme of the degree heterogeneity such as $\gamma < 3$, the negative degree-degree correlation causes the anti-correlation between $s(i)$ and $s(j)$~\cite{JSLee2006}, so the mutuality can actually be negative if the effect of such anticorrelation dominates [most notably, when the weight distribution is homogeneous---the upper left part of Fig.~\ref{fig:app_synthetic}(c)].

The correlation between LR values $\rho$ and AR values $\tau$ shows a more intricate behavior than the others, as depicted in Fig.~\ref{fig:app_synthetic}(d). One should first note that the correlation is determined by the correlation between the number of reciprocal edges emanating from a node in Eq.~\eqref{eq:rho} and the effective in-degree value of the node in Eq.~\eqref{eq:tau}, in the directed subnetwork. Large values of both $\gamma$ and $\lambda$ (homogeneous weight distributions on top of networks with homogeneous degree distributions---the upper right part of the plot) tend to conserve most of the original edges bidirectionally as in the case of AJD. In that case, there is not enough variability in the values of $\rho$ and $\tau$, both of which approach unity for most of the nodes, so the correlation itself becomes meaningless. As the values of $\gamma$ are decreased, however, the hub nodes with large values of degree dominate the system, and their effective in-degree and reciprocity start to increase simultaneously---they are positively correlated because large values of effective in-degree naturally increase the candidate for reciprocal edges [see Eqs.~\eqref{eq:rho} and \eqref{eq:tau}]. Similarly, the decreased value of $\lambda$ (heterogeneous weight distributions) increases the correlation by providing meaningful amount of variability in the values of $\rho$ and $\tau$ for different nodes, based on the same argument as in the small $\gamma$ values---large effective in-degree values mean large numbers of candidates for reciprocal edges.

Our scale-free model networks with uncorrelated power-law distributed weights admittedly lack many aspects of real-world networks, but we hereby provide the corresponding results for various measures presented in the main text as a guideline providing the baseline properties in unstructured networks. There can be different types of null-model networks and further analyses based on them, of course, and we believe that the results we present in this Appendix are the first step to figure out the landscape of the quantities of interest in the context of networks characterized by heavy-tailed distributions.

\section{Intricate relation between mutuality and reciprocity}
\label{sec:app_compare_m_r}

\begin{figure}[b]
\includegraphics[width=\columnwidth]{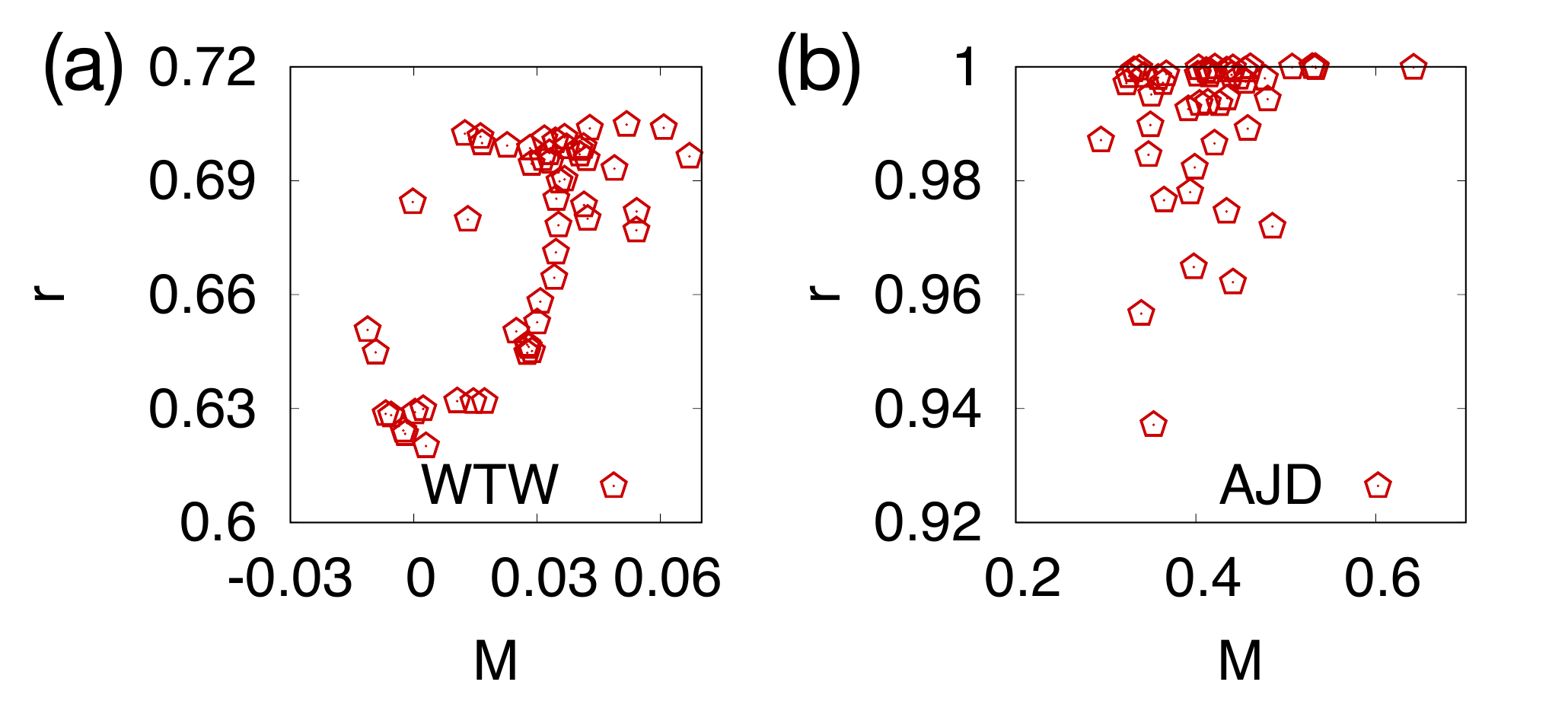}
\caption{Scatter plots representing the mutuality $M$ and the reciprocity $r$ in (a) the WTW and (b) the AJD, where each point corresponds to $(M,r)$ of each network.}
\label{fig:app_m_r}
\end{figure}

Both the mutuality $M$ and the reciprocity $r$ introduced in the main text are the measures to evaluate the mutual importance for a given network by utilizing the normalized weights as a starting point. However, they capture the concept of mutual importance in rather different perspectives. For an illustrative example for demonstration, let us consider a star network composed of $N$ nodes with $(N-1)$ edges that connect the central node $c$ to all of the other peripheral nodes. If all of the weights on the $(N-1)$ edges are the same, $\normw_{ci} = 1/(N-1)$ and $\normw_{ic} = 1$ for all of the peripheral nodes $i$. In this case, $M = -1$ because the normalized weights in the opposite direction are always anticorrelated (or completely disassortative~\cite{Assort2002,Assort2003}, as they are equivalent here), but $r = 1$ because the subnetwork retains all of the original edges bidirectionally due to the completely uniform weight values. If we look closely into the situation, we can see that the mutuality solely takes the bilateral relation, while the reciprocity contains the information on the overall weight distributions around each node used to extract the subnetwork. 

Therefore, although the mutuality may look intuitive and convenient to use (we need not calculate the entropy measures and others), it is not able to capture a more nuanced concept of mutual dependency: if we take the star network in the previous paragraph again, even if the central node is dominant in the structural aspect (captured by $M = -1$), all of the peripheral nodes are equally important to the central node as well (captured by $r = 1$). Of course, there are cases where the former is more relevant depending on the context, so we claim to use both measures to fully characterize a given networked system with weights. We show both measures for our data in Fig.~\ref{fig:app_m_r}, and one can observe that the relationship between two measures is not simple with quite scattered points.

\end{CJK}


\begin{thebibliography}{00}
\bibitem{ShannonEntropy}
By applying L'H{\^o}pital's rule, one can
easily show that $S_{\alpha \to 1} (i) = - \sum_{j \in \nu(i)} \tilde{w}_{ij} \ln \tilde{w}_{ij}$, which is the actually used formula to calculate the measure for the $\alpha = 1$ case in this work.
\end{thebibliography}
\end{document}